# Hierarchical Multiscale Modeling of Macromolecules and their Assemblies

P. Ortoleva[a], A. Singharoy[a], and S. Pankavich[b,c]

[a]Center for Cell and Virus Theory

Department of Chemistry

Indiana University

Bloomington, IN 47405

[b]Department of Mathematics
United States Naval Academy
Annapolis, MD 21402

[c]Department of Applied Mathematics and Statistics
Colorado School of Mines
Golden, CO 80401

Contact: pankavic@usna.edu

**Abstract**
Soft materials (e.g., enveloped viruses, liposomes, membranes and supercooled liquids) simultaneously deform or display collective behaviors, while undergoing atomic scale vibrations and collisions. While the multiple space-time character of such systems often makes traditional molecular dynamics simulation impractical, a multiscale approach has been presented that allows for long-time simulation with atomic detail based on the co-evolution of slowly-varying order parameters (OPs) with the quasi-equilibrium probability density of atomic configurations. However, this approach breaks down when the structural change is extreme, or when nearest-neighbor connectivity of atoms is not maintained. In the current study, a self-consistent approach is presented wherein OPs and a reference structure co-evolve slowly to yield long-time simulation for dynamical soft-matter phenomena such as structural transitions and self-assembly. The development begins with the Liouville equation for $N$ classical atoms and an ansatz on the form of the associated $N$-atom probability density. Multiscale techniques are used to derive Langevin equations for the coupled OP-configurational dynamics. The net result is a set of equations for the coupled stochastic dynamics of the OPs and centers of mass of the subsystems that constitute a soft material body. The theory is based on an all-atom methodology and an interatomic force field, and therefore enables calibration-free simulations of soft matter, such as macromolecular assemblies.

**I. Introduction**
A variety of soft matter systems can be viewed as an array of molecules transiently occupying lattice positions about which vibrational/rotational motion occurs. This suggests that such systems have mixed solid- and liquid-like behavior, and warrants a theory that naturally integrates the combined characteristics of these states. Examples of soft matter systems include macromolecular assemblies such as viruses and liposomes, liquid crystals, and super-cooled water. The objective of the present study is to develop a multiscale theory that (a) starts from the all-atom structure of the system; in particular, macromolecules or their assemblies, (b) introduces a set of coarse-grained variables for capturing their large-scale organization, and (c) arrives at a set of stochastic equations describing their long-time dynamics through the coupled evolution of these variables with a quasi-equilibrium ensemble of all-atom structures.

    Macromolecular assemblies exhibit a complex hierarchy of structural organization [1-2]. For example, viruses display a hierarchical organization of atoms forming protomers, pentamers or hexamers that ultimately assemble into capsids via different types of bonded and non-bonded interactions. This hierarchy results in the multiple space and timescale dependencies underlying pathways of structural organization, e.g., assemblies are organized on length scales from angstroms to tens of nanometers or more, and involve processes that occur on timescales ranging from femtoseconds to milliseconds. While molecular dynamics (MD) have been widely used to simulate macromolecular structures at an atomistic level, the simulation time for nanoscale assemblies has been limited to tens or sometimes a few hundreds of nanoseconds [3-4]. The constraint on the size of the timestep in MD does not allow for simulations to continue over long time periods of physical relevance, or for the simulation of systems with a large number of atoms. Feasibility of such simulations also depends on the extent of parallel computing resources available. Recently, billion-atom MD simulations have been accomplished [5-6]. However, they neglect one or more of the Coulomb interactions, bonded forces or rapidly fluctuating hydrogen atoms. All the latter are central to biomolecular structure and dynamics. Thus, all-atom simulation of large macromolecular assemblies has been a computational challenge.

    Interestingly, the very hierarchical nature of macromolecular assemblies has been utilized in designing reduced dimensionality frameworks to facilitate their efficient computer simulation, but is achieved at the expense of losing atomic scale resolution. This has resulted in the development of coarse-grained models such as bead [7], rigid-blob [8], shape-based [9], rigid region decomposition [10], symmetry-constrained [11] and curvilinear coordinate [12] models, as well as Principle Component Analysis (PCA) and Normal Mode Analysis (NMA) guided approaches [13-14]. Simulation methodologies based on these models involve tracking a much smaller number of dynamical variables than those based on all-atom description. Thus, the computational cost of implementing reduced dimensionality models is moderate. Similarly, a coarse-grained strategy has been introduced which uses Dissipative Particle Dynamics (DPD) simulations with an effective potential obtained by applying the inverse Monte Carlo (MC) method to an initial MD simulation [15-16]. Advantages and shortcomings of these approaches in the context of macromolecular simulations are reviewed in [17, 18] and references therein.

    In a series of recent studies, we have discovered novel multiscale techniques that probe the cross-talk (Fig. 1) among multiple scales in space and time that are inherent within such systems, yet preserve all-atom detail within the macromolecular assemblies [19-30]. This is achieved via the

introduction of a set of slowly-evolving Order Parameters (OPs) that describe coherent, overall structural changes. Broadly speaking, these OPs filter out the high frequency atomistic fluctuations from the low frequency coherent modes and have been utilized to capture a variety of effects of multiscale physical systems, including Ostwald ripening in nanocomposites [25], nucleation and front propagation pathways during a virus capsid structural transition [28], and counter-ion and temperature induced transitions in viral RNA [29-30]. As they evolve on a much longer time scale than that of atomistic processes, the OPs serve as the basis of a multiscale analysis. This analysis begins with the *N*-atom Liouville equation and yields rigorous Smoluchowski, Fokker-Planck, and Langevin equations for the stochastic dynamics of OPs. The resulting Langevin equations have been implemented via a force-field based algorithm in our Deductive Multiscale Simulator (DMS) software (previously called SimNanoWorld) and completely account for the dynamics of the OPs [17]. The theory is the natural consequence of a long history of multiscale analysis in classical many-particle physics [31-37].

The OPs, as introduced here, describe the global organization of many-particle systems and probe complex motions such as macromolecular twisting or bending. Classic examples of OPs include the degree of local preferred spin or molecular orientation, and mass density, for which profiles vary across a many-particle system. For a solid, profiles of particle deviation from rest lattice positions have traditionally been used. However, for soft matter the time scale of many phenomena is comparable to that for migration away from lattice positions, making the latter a less sensitive OP. Furthermore, classical phase transition theories, like that for magnetization, are built on the properties of infinite systems, e.g., renormalization group concepts [38]. In contrast, macromolecular assemblies are finite, in fact small in extent, and hence cannot completely follow the theory of macroscopic phase transitions. Also, macromolecular assemblies can reside in conformational states without a simple, readily identifiable symmetry, e.g., ribosomes. Nonetheless, as pH and other conditions in the host medium change, the system can switch to a different conformation [39]. Such a system experiences a structural transition between two states, neither of which has a readily identifiable symmetry. This suggests that often OPs in macromolecules cannot be readily associated with the breaking of symmetry even if they signify a dramatic change of order. Therefore, a new formulation is required to signal the emergence of new order in macromolecular systems when there are no readily identifiable symmetries involved.

For nanoscale assemblies OPs have been introduced as generalized mode amplitudes [40]. More precisely, vector OPs $\bar{\Phi}_k$ were constructed that characterize system dynamics as a deformation from a reference configuration of $N$ atoms $\{\bar{r}_i^0; i=1,\cdots,N\}$. The set of time-dependent atomic positions $\{\bar{r}_1,\cdots,\bar{r}_N\}$ was expressed in terms of a collection of basis functions $U_k(\bar{r}_i^0)$ and these OPs $\bar{\Phi}_k$ [19]. Thus, variations in the OPs generate the structural transformations. Since the OPs characterize overall deformation, the $U_k$ functions vary smoothly across the system, i.e., on the nanometer scale or greater. As one seeks only a few OPs (<<*N*), the above relationship between the atomic positions and OPs cannot completely describe individual atomic motions. This was previously addressed by introducing residuals to capture the short-scale atomic dynamics and deriving equations for the co-evolution of the OPs and probability distribution of atomic configurations [21-22, 24, 41-42].

However, as soft materials are easily deformed by thermal stress and fluctuation, large deformations of soft matter cannot be considered as coherent changes determined by merely a few OPs. To address this we introduce the notion that one might conceive of a slowly evolving hierarchal structure about which to formulate the construction of OPs. Instead of OPs depending explicitly upon the *N*-atom configuration, we introduce intermediary variables, representing the CMs of subsystems within the soft matter structure, and construct OPs which depend only upon these CMs. Additionally, rather than being constrained by an initial reference configuration, we reformulate the construction of basis functions to depend upon quantities that vary slowly with system-wide deformations. This allows for the methodology to accurately describe systems, such as macromolecular assemblies, which may undergo drastic deformations and hence cannot be modeled over long time periods as simple, continuous transformations from a fixed reference configuration. Demonstrated below, an example of such a situation arises when a system transitions from subsystem-disconnected to connected states, i.e., connected subsystems cannot be perceived as a simple deformation of disconnected ones. Additional advantages of using a slowly evolving reference configuration for defining the OPs are also discussed. Ultimately, a multiscale methodology based on these variables couples the atomistic and coarse-grained evolution to facilitate all-atom simulations of slow processes in complex assemblies. This theme is developed in the present article.

In subsequent sections we introduce new multiscale techniques to address the modeling and simulation of evolving soft-matter systems. First, intermediary subsystem centers of mass that characterize the meso-scale deformation of the system and OPs for tracking the long-scale migration of individual molecules are introduced (Sect. II). Multiscale techniques are then used to provide evolution equations for the OPs and subsystem center of mass variables (Sect. III). Next, a set of stochastic Langevin equations for the coupled dynamics of the OPs and CM variables are derived (Sect. IV). Finally, an algorithm suggested by the multiscale development is presented and validated via simulations of different structural components in satellite tobacco mosaic virus (STMV) (Sect. V). Conclusions are then drawn in Sect. VI.

## II. Collective Modes and Molecular Migration

In the present context, OPs are variables describing the long-scale organization of a system. In previous studies, OPs form the starting point of a multiscale theory of nanosystem dynamics [23, 25]. Since changes in the nanoscale features characterized by the OPs involve the coordinated motion of many atoms, OPs tend to evolve on time scales much longer than that of individual atomic vibration. This underlies the multiscale character of soft-matter.

Biological soft matter is typically organized in a hierarchical fashion. For example, a non-enveloped virus consists of about $N = 10^6$ atoms, organized into about $N^{sys} = 100$ macromolecules (e.g., protein and RNA or DNA). When the system is spheroidal (as for an icosahedral virus) it has a total diameter of about 100 typical atomic diameters, i.e., around $N^{1/3}$, while the total mass of the system is $N \cdot m$ where $m$ is the average atomic mass, and that of a typical macromolecule is about $mN / N^{sys}$. In order to accurately represent this multiple mass and length scale structure, we incorporate a hierarchal OP formulation into the description of the system.

On the finest scale, the dynamics of the physical system are described by the 6$N$ atomic positions and momenta, denoted by

$$\Gamma = \{\vec{r}_i, \vec{p}_i; i = 1, 2, \cdots, N\}. \tag{II.1}$$

Since the matter of interest is hierarchical by nature, we divide the overall structure into $N^{sys}$ non-overlapping subsystems indexed by $S = 1, 2, \ldots N^{sys}$. The center of mass of each subsystem, given by

$$\vec{R}^S = \sum_{i=1}^{N} \frac{m_i}{M^S} \vec{r}_i \Theta_i^S \tag{II.2}$$

will serve as an intermediate-scale description for the behavior of the material. Here, $m_i$ is the mass of atom $i$, $M^S$ is the mass of subsystem $S$, and the function $\Theta_i^S$ is one if atom $i$ is in subsystem $S$ and zero otherwise. Effectively, the $\vec{R}^S$ variables denote subsystem OPs that characterize the organization and dynamics of the $S^{th}$ subsystem.

While the centers of mass describe subsystem-wide motion, one must also describe the largest scale of interest to illustrate changes in the overall structure of the system. Thus, we introduce a set of hierarchical OPs $\underline{\Phi} = \{\bar{\Phi}_k; k = 1,2,...\}$ to further characterize collective behaviors. This is performed using a space-warping transformation [19] that is modified to accommodate the present hierarchical structure of soft matter. First, we introduce the relationship between OPs $\bar{\Phi}_k$ and CMs $\bar{R}^S$ by

$$\bar{R}^S = \sum_k \bar{\Phi}_k U_k^S(\underline{R}) \tag{II.3}$$

where $\underline{R} = \{\bar{R}^S; S = 1,2,...,N^{sys}\}$ is the set of all subsystem centers of mass and $U_k^S(\underline{R})$ is a pre-chosen basis function depending upon these CMs restricted to subsystem $S$. The basis function $U_k^S$ is constructed as $U_k(\bar{R}^S) = U_{k_1}(X^S)U_{k_2}(Y^S)U_{k_3}(Z_i^S)$, where $k$ is a set of three integers $k_1, k_2, k_3$ implying order of the Legendre polynomial $U$ for the $X, Y, Z$ components of $\bar{R}^S$ respectively. As in our previous work, OPs labeled by indices $k = \{000, 100, 010, 001\}$ are denoted lower-order and $k > \{000, 100, 010, 001\}$ are higher-order [43].

Notice that basis functions do not depend upon each atomic position $\bar{r}_i$, but rather on the intermediate scale variables, $\underline{R}$, thereby ensuring a hierarchal foundation. Additionally, since the basis functions $U_k^S$ depend on dynamic variables $\bar{R}^S$ (and not CMs of a fixed reference configuration, say $\bar{R}_0^S$), the collection of expressions (II.3) is an implicit equation for the CMs. By choosing the set of $U_k^S$ basis functions to be smoothly varying, the set of $\bar{\Phi}_k$ track the overall coherent deformation of the soft matter. Since coherent deformation of the entire structure implies slow motion, we expect that the $\bar{\Phi}_k$ variables will be slowly varying in comparison to both the migration of CMs and the fluctuation of atomic positions.

For a finite truncation of the sum in (II.3), there will be some residual displacements. Hence, (II.3) becomes

$$\bar{R}^S = \sum_k \bar{\Phi}_k U_k^S(\underline{R}) + \bar{\sigma}^S \tag{II.4}$$

where $\bar{\sigma}^S$ is the residual distance for the $S^{th}$ subsystem in the soft matter nanostructure. The OPs are then expressed precisely in terms of the $\bar{R}^S$ variables by minimizing the mass-weighted square residual

$$\Sigma = \sum_{S=1}^{N^{sys}} M^S |\bar{\sigma}^S|^2 \tag{II.5}$$

with respect to $\underline{\Phi}$ at constant $\underline{R}$. With (II.4), the expression for $\Sigma$ becomes

$$\Sigma = \sum_{S=1}^{N^{sys}} M^S \left| \bar{R}^S - \sum_k U_k^S(\underline{R}) \bar{\Phi}_k \right|^2. \tag{II.6}$$

The optimal OPs are those that minimize $\Sigma$, i.e. those containing the maximum amount of information so that the $\bar{\sigma}^S$ are on average the smallest. Thus, using (II.5) we obtain [24, 42] the system of equations:

$$\sum_{k'} B_{kk'} \bar{\Phi}_{k'} = \sum_{S=1}^{N^{sys}} M^S U_k^S(\underline{R}) \bar{R}^S \tag{II.7}$$

$$B_{kk'} = \sum_{S=1}^{N^{sys}} M^S U_k^S(\underline{R}) U_{k'}^S(\underline{R}). \tag{II.8}$$

If we choose a preliminary set of basis functions $U_k^S(\underline{R})$ to be, for instance, Legendre polynomials [21-

22, 41], then a Gram-Schmidt procedure can be used to generate an orthonormal basis. In particular, for the current study we simplify the formulation and choose the normalization such that

$$B_{kk'} = \mu_k \delta_{kk'} \tag{II.9}$$

$$\mu_k = \sum_{S=1}^{N^{sys}} M^S \{U_k^S(\underline{R})\}^2 \tag{II.10}$$

and thus obtain an explicit representation of the OPs

$$\vec{\Phi}_k = \frac{1}{\mu_k} \sum_{S=1}^{N^{sys}} M^S U_k^S(\underline{R}) \vec{R}^S \tag{II.11}$$

in terms of the subsystem CMs. As $\vec{\Phi}_k$ changes, space is deformed and so does the assembly embedded in it. Here $\mu_k$ serves as an effective mass associated with $\vec{\Phi}_k$ and is proportional to the square of the basis vector's length. The masses primarily decrease with increasing complexity of $U_k^S$ [30, 43]. Thus, the OPs with higher $k$ probe smaller regions in space. As the basis functions depend on the collection of CMs, (II.11) is an explicit equation for $\underline{\Phi}$ in terms of $\underline{R}$. Specific sets of OPs can capture deformations including extension, compression, rotation, tapering, twisting and bending. Such physically relevant interpretation of the OPs is provided in detail earlier [30], and in Sect. SI1 of the Supporting Information. Hence, we utilize three differing levels of description - the finest scale of atomic vibration captured by $\Gamma$, the set of intermediate scale CM variables for each subsystem $\underline{R}$, and a global set of slowly evolving OPs given by $\underline{\Phi}$.

To conclude the hierarchical construction of variables describing the system, we show that both the scaled subsystem CMs and global OPs may vary slowly relative to individual atomic fluctuations, and thus they can serve as the basis for a multiscale analysis (Sect. III). To reveal the respective time scales on which $\vec{\Phi}_k$ and $\vec{R}^S$ evolve, it is convenient to define smallness parameters, in this case $\varepsilon_1$, $\varepsilon_2$, and $\varepsilon_3$. Within the context of the current study, it is the ratio of masses and lengths that characterizes the significant difference in motion throughout the system. Since the subsystem mass is significantly larger than that of the average atom, the parameter $\varepsilon_1 = m/M^S$, where $m$ is a typical atomic mass will accurately describe this separation of scales. In a similar manner, as $\mu_k$ represents the sum of subsystem masses we assume that this quantity is large with respect to $M^S$. Hence, we define $\varepsilon_2 = M^S/M_{TOT}$ where $M_{TOT}$ is the mass of the entire system, and write $\mu_k \approx M_{TOT} = M^S/\varepsilon_2 = m/\varepsilon_1\varepsilon_2$ for the effective mass related to the $k$th OP. Finally, the length scales of interest may also differ greatly. Thus, one may define another smallness parameter $\varepsilon_3 = L^S/L^{sys}$ where $L^S$ represents the average nearest-neighbor subsystem difference and $L^{sys}$ is the radius of the entire system. The effect of this length scale difference enters into the analysis through the use of the basis functions. Namely, as the basis functions vary slowly across the entire system, they may depend not upon subsystem CMs, but instead on scaled CM variables $\tilde{R} = \varepsilon_3 \bar{R}$. However, for the purposes of the current study, we will assume that the difference in length scales is not sufficiently large in comparison to the mass scalings, and hence $\varepsilon_3 = O(1)$.

To investigate the time rate of change of $\vec{\Phi}_k$ and $\vec{R}^S$, we make use of the Liouville operator

$$L = -\sum_{i=1}^{N} \left( \frac{\vec{p}_i}{m_i} \cdot \frac{\partial}{\partial \vec{r}_i} + \vec{F}_i \cdot \frac{\partial}{\partial \vec{p}_i} \right) \tag{II.12}$$

where $\vec{p}_i$ and $\vec{F}_i$ represent the momentum of, and the net force acting on, atom $i$ respectively. Using

(II.11) we compute $\frac{d\vec{\Phi}_k}{dt} = L\vec{\Phi}_k$ and $\frac{d\vec{R}^S}{dt} = L\vec{R}^S$ to find

$$\frac{d\vec{\Phi}_k}{dt} = \frac{1}{\mu_k}\left(\vec{\Pi}_k + \vec{\pi}_k\right) \quad \text{(II.13)}$$

$$\frac{d\vec{R}^S}{dt} = \frac{\vec{P}^S}{M^S} \quad \text{(II.14)}$$

where $\vec{P}^S = \sum_i \vec{p}_i \Theta_i^S$ is the total momentum of the $S^{\text{th}}$ subsystem. Additionally, the terms appearing in (II.13) are given by

$$\vec{\Pi}_k = \sum_{S=1}^{N^{sys}} U_k^S(R)\vec{P}^S \quad \text{(II.15)}$$

$$\vec{\pi}_k = \sum_{i=1}^{N} \frac{\partial U_k^{S(i)}}{\partial R^{S(i)}} \vec{p}_i \cdot \vec{R}^{S(i)} \quad \text{(II.16)}$$

Here it is seen that $\vec{\Pi}_k$ is the conjugate momentum associated with the $k^{\text{th}}$ OP $\vec{\Phi}_k$, while $\vec{\pi}_k$ appears due to the dependence of basis functions $U_k^S$ on intermediate-scale CMs. Using the definitions of the smallness parameters, (II.13) and (II.14) yield

$$\frac{d\vec{\Phi}_k}{dt} = \varepsilon_1 \varepsilon_2 \frac{\vec{\Pi}_k + \vec{\pi}_k}{m} \quad \text{(II.17)}$$

$$\frac{d\vec{R}^S}{dt} = \varepsilon_1 \frac{\vec{P}^S}{m} \quad \text{(II.18)}$$

for any of the OPs or CMs. For different systems, and thus different choices of $\varepsilon_1$ and $\varepsilon_2$, these variables may evolve slowly relative to atomic vibration. For instance, if one considers a case in which the motion of the subsystems is of greater interest than that of the overall structure, then the ratio of subsystem mass to total mass is much smaller than that of atomic mass to subsystem mass (i.e., $\varepsilon_2 \ll \varepsilon_1 \ll 1$). Within this framework, one may view the CMs as rapidly fluctuating with respect to the time scale of motion of the OPs. Contrastingly, if the mass ratio relationship is reversed (i.e., $\varepsilon_1 \ll \varepsilon_2 \ll 1$), then the CMs and OPs may evolve on similar time scales. For the purposes of the current study, we consider the situation in which the total system consists of a relatively small number of subsystems (e.g., a few pentamers), and hence the latter formulation applies. Therefore, the second smallness parameter remains large relative to the first, i.e. $\varepsilon_2 = O(1)$. Additionally, we rewrite the first parameter as $\varepsilon_1 = \varepsilon$ for $\varepsilon$ small. From this, (II.17) and (II.18) demonstrate that the CMs and OPs evolve slowly, at a rate $O(\varepsilon)$, in relation to the atomistic variables. Thus, our formulation is consistent with the quasi-equilibrium distribution of all-atom configurations $\Gamma$ at fixed values of $\vec{R}^S$ and $\vec{\Phi}_k$. We conclude that the set of $\vec{\Phi}_k$ and $\vec{R}^S$ describe slow dynamics of the soft matter. The OPs $\vec{\Phi}_k$ describe overall deformational (collective) behavior while the long scale motion of the macromolecular CMs is described via the $\vec{R}^S$. Other variables such as the preferred orientation of the macromolecules could also be included [44], but for simplicity are not within the present study.

In subsequent sections, we use the slow behavior of these variables as one element of the multiscale development. Additionally, we show that the pair $\left(\vec{R}^S, \vec{\Phi}_k\right)$ satisfy Langevin dynamics in contrast with the $\vec{r}_i$ which do not, due to the key role of inertial effects underlying the vibrational motion of individual atoms and the long scale nature of the CM dynamics tracked by $\vec{R}^S$ and $\vec{\Phi}_k$.

### III. Multiscale Analysis

The present theory of soft matter is based on the notion that there are multiple distinct time scales on which system evolution takes place. The shortest time scale is that of atomic vibrational motion about slowly changing reference positions. The slow dynamics is attributed to intermediate scale molecular migration ($\underline{R}$ the set of macromolecular CMs) and long-scale, coherent deformation described by OPs $\vec{\Phi}_k$. Thus, a multiscale approach is a natural solution to the challenges presented by the operation of many processes across various scales in space and time.

To begin the multiscale analysis we first use the Liouville equation to derive a conservation law for the slow dynamics of the system. We define $\tilde{W}$ to be the joint probability density for $\underline{\Phi}$ and $\underline{R}$:

$$\tilde{W}(\underline{\Phi},\underline{R},t) = \int d\Gamma^* \Delta(\underline{\Phi}-\underline{\Phi}^*)\Delta(\underline{R}-\underline{R}^*)Y(\Gamma^*,t) \tag{III.1}$$

where the superscript * indicates evaluation at $\Gamma^*$, the $N$-atom position-momentum state over which integration is taken. Note that simultaneous introduction of $\underline{\Phi}$ and $\underline{R}$ does not imply a contradiction. This follows from the fact that $\underline{R}$ is the set of center-of-masses of peptides, nucleotides, or other polyatomic substructure. Using the smallness parameter from Sect. II, we arrive at the following representation for derivatives of CMs and OPs:

$$\frac{d\vec{\Phi}_k}{dt} = \varepsilon\frac{\vec{\Pi}_k + \vec{\pi}_k}{m} \tag{III.2}$$

$$\frac{d\vec{R}^S}{dt} = \varepsilon\frac{\vec{P}^S}{m}. \tag{III.3}$$

With this we use the Liouville equation for the $N$-atom probability density $Y(t,\Gamma)$

$$\partial Y/\partial t = LY \tag{III.4}$$

with Liouville operator given by (II.12) and arrive at a conservation law [23] for $\tilde{W}$ via the chain rule. Namely, we find

$$\frac{\partial \tilde{W}}{\partial t} = -\varepsilon \int d\Gamma^* \left[\sum_k \frac{\vec{\Pi}_k^* + \vec{\pi}_k^*}{m} \cdot \frac{\partial}{\partial \vec{\Phi}_k} + \sum_{i=1}^N \frac{\vec{P}^{S(i)*}}{m} \cdot \frac{\partial}{\partial \vec{R}^{S(i)}}\right] \Delta(\Gamma^*,\underline{\Phi},\underline{R})Y \tag{III.5}$$

where $\Delta(\Gamma^*,\underline{\Phi},\underline{R}) = \Delta(\underline{\Phi}-\underline{\Phi}^*)\Delta(\underline{R}-\underline{R}^*)$. This equation involves $Y$ and is thus not closed with respect to $\tilde{W}$. However, one finds [23] that this formulation enables a novel procedure for constructing a closed equation for $\tilde{W}$ when $\varepsilon$ is small. The expressions for $\vec{\Pi}_k$ and $\vec{\pi}_k$ can be determined explicitly by (II.15) and (II.16).

Throughout, we will make use of the hypothesis that the $N$-atom probability density $Y$ for the soft-matter assembly has multiscale character. Thus, $Y$ can be represented to express its dependence on the atomic positions and momenta (denoted collectively by $\Gamma$) both directly and, via the set of OPs $\underline{\Phi}$ and centers of mass $\underline{R}$ introduced in Sect. II, indirectly. To reflect the multiple time scale character of the system, the $N$-atom probability density $Y$ is written with multiple dependencies on atomic and collective variables in the form

$$Y = \rho(\Gamma,\underline{\Phi},\underline{R};t_0,\underline{t};\varepsilon). \tag{III.6}$$

The time variables $t_n = \varepsilon^n t$, are introduced to track processes on time scales $O(\varepsilon^{-n})$ for $n = 0,1,2,\cdots$.

The set $\underline{t} = \{t_1, t_2, \cdots\}$ tracks time for the slow processes, i.e., much slower than those on the $10^{-14}$ second scale of atomic vibrations. In contrast, $t_0$ tracks the latter fast atomistic processes. The ansatz on the dependence of $Y$ is not a violation of the number $(6N)$ of degrees of freedom, but rather a way to express the multiple ways in which $Y$ depends on $\Gamma$ and $t$.

With this, the ansatz and chain rule imply that the Liouville equation takes the form

$$\sum_{n=0}^{\infty} \varepsilon^n \frac{\partial \rho}{\partial t_n} = (L_0 + \varepsilon L_1) \rho, \qquad (\text{III.7})$$

$$L_0 = -\sum_{i=1}^{N} \left( \frac{\vec{p}_i}{m_i} \cdot \frac{\partial}{\partial \vec{r}_i} + \vec{F}_i \cdot \frac{\partial}{\partial \vec{p}_i} \right), \qquad (\text{III.8})$$

$$L_1 = -\sum_{k} \frac{\vec{\Pi}_k + \vec{\pi}_k}{m} \cdot \frac{\partial}{\partial \vec{\Phi}_k} - \sum_{i=1}^{N} \frac{\vec{P}^{S(i)}}{m} \cdot \frac{\partial}{\partial \vec{R}^{S(i)}}, \qquad (\text{III.9})$$

The operator $L_1$ involves partial derivatives with respect to $\underline{\Phi}$ and $\underline{R}$ at constant $\Gamma$, whereas, the converse is true for $L_0$, which involves partial derivatives with respect to $\Gamma$ at constant values of $\underline{\Phi}$ and $\underline{R}$. By mapping the Liouville problem to a higher dimensional description (i.e., from $\Gamma$ to $\underline{\Phi}, \underline{R}, \Gamma$) our strategy [23-24] is to solve the equation perturbatively in the higher dimensional representation and the small $\varepsilon$ limit. Using this perturbative approximation for $\rho$ and the exact conservation law (III.5), we obtain a closed equation for $\tilde{W}$. Since $\varepsilon$ is small, the development can be advanced with an expansion

$$\rho = \sum_{n=0}^{\infty} \varepsilon^n \rho_n. \qquad (\text{III.10})$$

Next, the multiscale Liouville equation is examined to each order in $\varepsilon$. We seek the lowest order behavior of $\rho$ which is slowly varying in time since the phenomena of interest take place on the microsecond or longer time scale, and not the $10^{-14}$ second scale tracked by $t_0$. Thus we assume the lowest order solution $\rho_0$ is independent of $t_0$, and validate this assumption by demonstrating the self-consistency of the solution that is constructed. This implies $\underline{\Phi}$ and $\underline{R}$ evolve slowly while the atomic variables $\Gamma$ come to a quasi-equilibrium state that co-evolves with them. To lowest order in the multiscale perturbation scheme, one obtains $L_0 \rho_0 = 0$. Using an entropy maximization procedure [24] with the canonical constraint of fixed average energy, the lowest order solution is determined to be

$$\rho_0 = \frac{e^{-\beta H}}{Q(\underline{\Phi}, \underline{R})} W(\underline{\Phi}, \underline{R}, \underline{t}) \equiv \hat{W}, \qquad (\text{III.11})$$

where $\beta$ is inverse temperature, $H$ is the Hamiltonian

$$H(\Gamma) = \sum_{i=1}^{N} \frac{p_i^2}{2m_i} + V(\vec{r}_1, \ldots, \vec{r}_n) \qquad (\text{III.12})$$

for $N$-atom potential $V$ and $\underline{\Phi}, \underline{R}$-dependent partition function $Q = \int d\Gamma^* \Delta(\Gamma^*, \underline{\Phi}, \underline{R}) e^{-\beta H^*}$.

To $O(\varepsilon)$ one obtains

$$\frac{\partial \rho_1}{\partial t_0} - L_0 \rho_1 = -\left( \frac{\partial \rho_0}{\partial t_1} - L_1 \rho_0 \right). \qquad (\text{III.13})$$

This equation admits the solution

$$\rho_1 = -\int_{-t_0}^{0} ds\, e^{-L_0 s}\left[\frac{\partial \rho_0}{\partial t_1} - L_1 \rho_0\right] \tag{III.14}$$

where the initial value of the first-order distribution is taken to be zero (i.e., the system is in a quasi-equilibrium state at the initial time). Using (III.9) and (III.11) in (III.14), one finds

$$\rho_1 = -t_0 \hat{\rho}\frac{\partial W}{\partial t_1} - \int_{-t_0}^{0} e^{-L_0 s}\hat{\rho}\left[\sum_k \frac{\vec{\Pi}_k + \vec{\pi}_k}{m}\left(\beta \vec{f}_k^{\Phi} W - \frac{\partial W}{\partial \vec{\Phi}_k}\right) + \sum_{i=1}^{N} \frac{\vec{P}^{S(i)}}{m}\left(\beta \vec{f}_i^{R} W - \frac{\partial W}{\partial \vec{R}^{S(i)}}\right)\right] \tag{III.15}$$

$$\vec{f}_k^{\Phi} = -\frac{\partial F}{\partial \vec{\Phi}_k},\quad \vec{f}_i^{R} = -\frac{\partial F}{\partial \vec{R}^{S(i)}} \tag{III.16}$$

for $(\vec{R}^S, \vec{\Phi}_k)$-constrained Helmholtz free energy $F$ where $Q = e^{-\beta F}$. Using the Gibbs Hypothesis, which states the equivalence of long-time and thermal ($\hat{\rho}$-weighted) averages, we find

$$\lim_{t_0 \to \infty} \frac{1}{t_0}\int_{-t_0}^{0} ds\, e^{-L_0 s} A = \int d\Gamma \hat{\rho} A \equiv A^{th} \tag{III.17}$$

for any dynamical variable $A$. As the $\vec{\Pi}_k$ involve the sum of momentum variables which tend to cancel, their thermal averages are zero and hence $\vec{\Pi}_k^{th} = 0$. Using this, dividing by $t_0$ in (III.15), and taking the limit as $t_0$ approaches infinity, we find $\lim_{t_0 \to \infty}\frac{\rho_1}{t_0} = -\frac{\partial W}{\partial t_1}$. Thus, removing divergent behavior as $t_0 \to \infty$ implies $\partial W/\partial t_1 = 0$ and hence $W$ is independent of $t_1$. Therefore, (III.15) becomes

$$\rho_1 = -\int_{-t_0}^{0} e^{-L_0 s}\hat{\rho}\left[\sum_k \frac{\vec{\Pi}_k + \vec{\pi}_k}{m}\left(\beta \vec{f}_k^{\Phi} W - \frac{\partial W}{\partial \vec{\Phi}_k}\right) + \sum_{i=1}^{N} \frac{\vec{P}^{S(i)}}{m}\left(\beta \vec{f}_i^{R} W - \frac{\partial W}{\partial \vec{R}^{S(i)}}\right)\right] \tag{III.18}$$

As we seek a kinetic theory correct to $O(\varepsilon^2)$, we can approximate $\rho$ via $\rho \approx \rho_0 + \varepsilon\rho_1$. Using the conservation law (III.5), along with (III.11) and (III.18) we arrive at a closed equation for $W$. Since $\tilde{W} \to W$ as $\varepsilon \to 0$, one obtains

$$\frac{\partial W}{\partial \tau} = \left\{\sum_{i,j=1}^{N}\frac{\partial}{\partial \vec{R}^{S(i)}}\cdot\left[\vec{\vec{D}}_{ij}^{RR}\left(\frac{\partial}{\partial \vec{R}^{S(j)}} - \beta \vec{f}_j^{R}\right)\right] + \sum_{k,k'}\frac{\partial}{\partial \vec{\Phi}_k}\cdot\left[\vec{\vec{D}}_{kk'}^{\Phi\Phi}\left(\frac{\partial}{\partial \vec{\Phi}_{k'}} - \beta \vec{f}_{k'}^{\Phi}\right)\right]\right.$$

$$\left.+\sum_{i=1}^{N}\sum_k \frac{\partial}{\partial \vec{R}_i}\cdot\left[\vec{\vec{D}}_{ik}^{\Phi R}\left(\frac{\partial}{\partial \vec{\Phi}_k} - \beta \vec{f}_k^{\Phi}\right)\right] + \sum_{i=1}^{N}\sum_k \frac{\partial}{\partial \vec{\Phi}_k}\cdot\left[\vec{\vec{D}}_{ik}^{R\Phi}\left(\frac{\partial}{\partial \vec{R}^{S(i)}} - \beta \vec{f}_i^{R}\right)\right]\right\}W \tag{III.19}$$

where $\tau = \varepsilon^2 t$. The diffusion ($D$) coefficients are given in terms of correlation function expressions

$$\vec{\vec{D}}_{ij}^{RR} = \frac{1}{m^2}\int_{-\infty}^{0} dt\left\langle e^{-L_0 t}\vec{P}^{S(i)}\cdot\vec{P}^{S(j)}\right\rangle \tag{III.20}$$

$$\vec{\vec{D}}_{kk'}^{\Phi\Phi} = \frac{1}{m^2}\int_{-\infty}^{0} dt\left\langle e^{-L_0 t}\left(\vec{\Pi}_k + \vec{\pi}_k\right)\cdot\left(\vec{\Pi}_{k'} + \vec{\pi}_{k'}\right)\right\rangle \tag{III.21}$$

$$\vec{\vec{D}}_{ik}^{\Phi R} = \frac{1}{m^2}\int_{-\infty}^{0} dt\left\langle e^{-L_0 t}\left(\vec{\Pi}_k + \vec{\pi}_k\right)\cdot\vec{P}^{S(i)}\right\rangle \tag{III.22}$$

$$\bar{\bar{D}}_{ik}^{R\Phi} = \frac{1}{m^2}\int_{-\infty}^{0}dt\left\langle e^{-L_0 t}\vec{P}^{S(i)}\cdot\left(\vec{\Pi}_k + \vec{\pi}_k\right)\right\rangle \tag{III.23}$$

where $\langle\cdots\rangle$ represents a thermal average over the $\underline{\Phi},\underline{R}$-constrained ensemble. The above equation for $W$ is of Smoluchowski form, and describes the evolution of the reduced probability density depending upon a set of CMs $\underline{R}$ evolving and interacting with a set of collective variables $\underline{\Phi}$. Though our analysis has obtained this equation up to $O(\varepsilon^2)$, the multiscale expansion can be extended to higher order of $\varepsilon$ to obtain an augmented Smoluchowski-type equation [44]. On the timescale on which the correlation functions decay for the present problem, the OPs are essentially unchanged. Therefore, to a very good approximation, the evolution in the correlation function occurs at constant OP values. This is simple to implement as correlation functions can then be computed via standard MD codes (Sect. V). This is distinct from approaches wherein the role of atomic scale fluctuation is not completely accounted for [45]. In this way, the atomic-nanoscale feedback loop of Fig. 1 in which the OPs affect the atomistic variables (and vice versa) is fully addressed. In the next section, we will derive the associated Langevin equations that describe the stochastic dynamics of $\underline{R}$ and $\underline{\Phi}$, and provide a computational foundation from which we can simulate the behavior of the intermediate and collective variables.

## IV. Langevin Equations for Soft Matter and Mutiscale Simulation Algorithm

The Smoluchowski equation provides a sound theoretical framework for stochastic OP dynamics. For practical computer simulation of viral systems, rigorous Langevin equations for the OPs equivalent to the above Smoluchowski equation can be derived [24-26]. In an analogous fashion, this is done below to arrive at stochastic equations for the simulation of soft matter. Though we will construct a precise mathematical argument, a more physically-rooted argument can also be utilized [23, 25]. We first group the $k_{max}$ OPs $\underline{\Phi}$ and $N^{sys}$ centers of mass $\underline{R}$ into a set of $M = k_{max} + N^{sys}$ coupled variables, represented by $\underline{\Psi} = \{\vec{\Psi}_k; k=1,...,M\}$. Then, we can rewrite the Smoluchowski equation (III.19) in terms of the unified OP representation as

$$\frac{\partial W}{\partial \tau} = \sum_{k,k'=1}^{M}\frac{\partial}{\partial\vec{\Psi}_k}\cdot\left[\bar{\bar{D}}_{kk'}\left(\frac{\partial}{\partial\vec{\Psi}_{k'}}+\vec{f}_{k'}\right)\right]W \tag{IV.1}$$

where the new diffusion coefficients $\bar{\bar{D}}_{kk'}$ are defined in terms of the diffusion coefficients of (III.20-III.23). With this representation of the equation for $W$, we will demonstrate that the associated Langevin equations take the form

$$\frac{d\vec{\Psi}_k}{dt} = \sum_{k'=1}^{M}\bar{\bar{\gamma}}_{kk'}\left\langle\vec{f}_{k'}(\underline{\Psi})\right\rangle + \vec{\xi}_k(t) \tag{IV.2}$$

for $k=1,...,M$ where $\langle\vec{f}_k\rangle$ is the thermal-averaged force, $\bar{\bar{\gamma}}_{kk'}$ is a coefficient matrix related to $\bar{\bar{D}}_{kk'}$ and $\vec{\xi}_k$ is a random force. Here, we assume the process generating $\vec{\xi}_k(t)$ is stationary and all of the average random forces vanish; hence $\xi_k$ satisfies for every $k=1,..,M$

$$\left\langle\vec{\xi}_k\right\rangle = 0, \tag{IV.3}$$

$$\left\langle\vec{\xi}_k(t)\vec{\xi}_{k'}(t')\right\rangle = \bar{\bar{A}}_{kk'}(t-t') \tag{IV.4}$$

where $\bar{\bar{A}}_{kk'}$ is a function to be determined. We denote the product of $M$ delta functions, one for each coupled OP by

$$\Delta(\underline{\psi},t) = \delta(\bar{\psi}_1 - \bar{\Psi}_1(t)) \cdots \delta(\bar{\psi}_M - \bar{\Psi}_M(t)) \tag{IV.5}$$

The $\Delta$-weighted probability density for these random variables is

$$W(\underline{\psi},t) = \langle \Delta(\underline{\psi},t) \rangle. \tag{IV.6}$$

Taking the time derivative of this equation we find

$$\frac{\partial W}{\partial t} = \left\langle \frac{\partial \Delta}{\partial t} \right\rangle. \tag{IV.7}$$

As the dynamics of $\bar{\Psi}_k$ are driven by a stochastic differential equation, the standard chain rule for real-valued functions does not apply; hence more care must be taken in evaluating the derivative of $\Delta$ in (IV.7). Using a stochastic version of the chain rule, known as Ito's Lemma [46-47], we find

$$\frac{\partial \Delta}{\partial t} = -\sum_{k=1}^{M} \frac{\partial \Delta}{\partial \bar{\psi}_k} \cdot \frac{d\bar{\Psi}_k}{dt} + \frac{1}{2} \sum_{k,k'=1}^{M} \frac{\partial^2 \Delta}{\partial \bar{\psi}_k \partial \bar{\psi}_{k'}} \cdot \frac{d[\bar{\Psi}_k, \bar{\Psi}_{k'}]}{dt} \tag{IV.8}$$

where $[\bar{\Psi}_k, \bar{\Psi}_{k'}]$ is the $m \times m$ quadratic covariation matrix generated by $\underline{\Psi}$. Using this in (IV.7) yields

$$\frac{\partial W}{\partial t} = -\sum_{k=1}^{M} \frac{\partial}{\partial \bar{\psi}_k} \cdot \left\langle \Delta \frac{d\bar{\Psi}_k}{dt} \right\rangle + \frac{1}{2} \sum_{k,k'=1}^{M} \frac{\partial^2}{\partial \bar{\psi}_k \partial \bar{\psi}_{k'}} \cdot \left\langle \Delta \frac{d[\bar{\Psi}_k, \bar{\Psi}_{k'}]}{dt} \right\rangle \tag{IV.9}$$

Integrating (IV.2) and using (IV.3)-(IV.4), we find

$$\left\langle \Delta \frac{d[\bar{\Psi}_k, \bar{\Psi}_{k'}]}{dt} \right\rangle \approx \left\langle \Delta \int_{-\infty}^{0} dt \bar{\bar{A}}_{kk'}(t) \right\rangle \equiv \langle \Delta \bar{\bar{\tau}}_{kk'} \rangle \tag{IV.10}$$

for every $k, k' = 1,\ldots,M$. Using (IV.2) and the independence of $W$ with the covariation, (IV.9) becomes

$$\frac{\partial W}{\partial t} = -\sum_{k=1}^{M} \frac{\partial}{\partial \bar{\psi}_k} \cdot \left[ \left( \sum_{k'=1}^{M} \bar{\bar{\gamma}}_{kk'} \langle \bar{f}_{k'}(\underline{\Psi}) \rangle + \langle \bar{\xi}_k \rangle \right) W \right] + \frac{1}{2} \sum_{k,k'=1}^{M} \frac{\partial^2}{\partial \bar{\psi}_k \partial \bar{\psi}_{k'}} \cdot [\bar{\bar{\tau}}_{kk'} W]. \tag{IV.11}$$

Again using (IV.3) and bringing the $k'$ sum out with the $k$ sum, this becomes a Smoluchowski equation:

$$\frac{\partial W}{\partial t} = -\sum_{k,k'=1}^{M} \frac{\partial}{\partial \bar{\psi}_k} \cdot \left[ \bar{\bar{\gamma}}_{kk'} \langle \bar{f}_{k'}(\underline{\Psi}) \rangle - \frac{\bar{\bar{\tau}}_{kk'}}{2} \frac{\partial}{\partial \bar{\psi}_{k'}} \right] W. \tag{IV.12}$$

Hence, we see that the solution of (IV.1), $W$, must be the probability density corresponding to the stochastic process $\underline{\Psi}$ that satisfies the associated Langevin equation (IV.2).

At long times the closed system reaches equilibrium. Hence

$$\lim_{t \to \infty} W(\underline{\Psi},t) = \frac{\exp(-\beta F(\underline{\Psi}))}{Z} \equiv W^{eq}(\underline{\Psi}), \tag{IV.13}$$

where $Z = \int d\Gamma^* e^{-\beta H^*}$. Recall that the first order correction $\rho_1$ has the form (III.18). After removal of the secular term in (III.15), all other terms vanish for $W = W^{eq}$ as can be verified algebraically. Thus, as $t \to \infty$, $\rho_1$ tends to zero, and hence using (III.10-11) and (IV.13), $\rho \approx \rho_0 = \hat{\rho} W^{eq} = \frac{e^{-\beta H}}{Z}$. Therefore, the long time limit of the system is the normalized Boltzmann distribution, which further implies that the classical equipartition theorem is satisfied.

As $W^{eq}$ must be a time-independent solution of (IV.12), we find

$$\bar{\bar{\gamma}}_{kk'} = \frac{\beta \bar{\bar{\tau}}_{kk'}}{2} = \frac{\beta}{2} \int_{-\infty}^{0} dt \bar{\bar{A}}_{kk'}(t) \tag{IV.14}$$

for all $k, k' = 1,\ldots,M$. With this, comparison of the Smoluchowski equation (IV.1) with the newly-derived formula (IV.12) implies the relation

$$\beta \vec{\vec{D}}_{kk'} = \vec{\vec{\gamma}}_{kk'} \tag{IV.15}$$

for all $k, k' = 1,\ldots,M$. Thus, the friction coefficients $\vec{\vec{\gamma}}_{kk'}$, and therefore the matrix $\vec{\vec{A}}_{kk'}$, are related to the diffusion coefficients $\vec{\vec{D}}_{kk'}$.

A commonly used approach for treating far-from-equilibrium systems involves projection operators [48]. It is very general in the sense that no approximations are made in arriving at an equation for the reduced probability of a subset of variables (OPs in our case) [49]. However, this kinetic equation requires the representation of a memory kernel, which usually can only be constructed using extensive MD simulations or experimental data. This is numerically expensive for *N*-atom problems except when the memory functions have short relaxation times [50]. In our analysis, the OPs of interest are much slower than the characteristic rate of atomistic fluctuations, and therefore the relaxation times are typically short relative to characteristic times of OP dynamics [29]. Under these conditions, our multiscale approach leads to the same set of Langevin equations as those from projection operators. However, the multiscale approach is more direct - we do not start with the projection operators, utilize a Markov assumption, or eventually resort to perturbation methods for approximating memory functions. Rather we make an *ansatz* that the *N*-atom probability density has multiple (initially unspecified) space-time dependencies, and analyze the resulting Liouville equation.

The OP and CM velocity autocorrelation function provides a criterion for the applicability of the present multiscale approach. If the reduced description is complete, i.e., the set of OPs and CMs considered do not couple strongly with other slow variables, then the correlation functions decay on a time scale much shorter than the characteristic time(s) of OP evolution [29]. However, if some slow modes are not included in the set of OPs, then these correlation functions can decay on timescales comparable to those of OP dynamics [29, 51]. This is because the missing slow modes, now expressed through the all-atom dynamics, couple with the adopted set of OPs. The present approach fails under such conditions. For example, setting the lower limit of the integral in (III.20-23) to $-\infty$ may fail to be a good approximation and the decay might not be exponential; rather it may be extremely slow so that the diffusion factor diverges. Consequently, atomistic ensembles required to capture such long-time tail behavior in correlation functions are much larger than those for capturing a rapid decay. Here, such situations are avoided via an automated procedure of understanding the completeness of the reduced description and adding additional OPs when needed (as discussed in Sect. SI2, and Refs. [29-30]). Adapting this strategy ensures that the OP velocity autocorrelation functions decay on timescales that are orders of magnitude shorter than those characterizing coherent OP dynamics, and thus the present multiscale approach applies. Next, we will develop a simulation algorithm to utilize this formulation of the problem and its description in terms of the Langevin equations.

Our starting point for the multiscale algorithm is the deformation of the initial reference configuration discussed in detail in Sect. II and the Langevin equations (IV.2). Given the all-atom structure of a macromolecular assembly at time $t=0$, $N^{sys}$ number of subsystems is identified and their CMs ($\vec{R}^S$) are calculated. These subsystem CMs are then used to construct the global OPs $\vec{\Phi}_k$ (II.11). Thus, the structural hierarchy of an assembly is captured. Then, multiple short MD simulations are used to construct a quasi-equilibrium ensemble of atomic configurations consistent with the instantaneous $\underline{\Phi}$ and $\underline{R}$ description [29, 51]. This ensemble of all-atom structures is employed to construct the diffusion factors, $\vec{\vec{D}}_{kk'}$ (III.20-23, IV.15), and thermal-averaged forces, $\langle \vec{f}_k \rangle$ (III.16). Details of the ensemble generation procedure and construction of these factors are provided in Supporting Information, and presented earlier in Appendix B of Ref. [43]. Using these forces and diffusions, the OPs and subsystem CMs are evolved via the Langevin equation to capture overall assembly deformations. As these equations form a coupled system of $M = k_{max} + N^{sys}$ stochastic differential equations, the dimensionality of the problem is reduced since the number of scaled molecular CM positions and OPs is much less than the total number of atoms in the system. Updating the set of CMs every Langevin timestep enables the reference configuration to slowly vary with the system over long times. The updated reference configuration $\underline{R}$ is used to compute new basis set polynomials. Using these $U_k^S$ and the Langevin evolved $\bar{\underline{\Phi}}_k$, an ensemble of CM configurations is constructed via (II.4). Each member in this ensemble is consistent with the instantaneous state of $\underline{\Phi}$. With this, the reference configuration, OPs and OP constrained ensemble of CMs are simultaneously evolved. Since the evolution of the OPs is inherently connected to that of the scaled positions and these are, in turn, dependent upon atomic trajectories, the algorithm should be completed by a procedure that allows repositioning of the atoms consistent with the overall structure provided by $\underline{R}$ and $\underline{\Phi}$. With the new set of atomic positions, forces and diffusions are recalculated to enable further Langevin evolution of the system. Thus, OPs constrain the ensemble of subsystem and atomic states (II.1-4), while the latter determine the diffusion factors (III.20-23) and thermal-average forces (III.16) that control OP evolution ((IV.2)). In this way, the ensemble of atomic configurations coevolves with the global OPs and subsystem CMs.

Water and ions are accounted for via the quasi-equilibrium ensemble (i.e., the configuration of the water and ions rapidly explores a quasi-equilibrium ensemble at each stage of the OP dynamics). This assumption holds only when water/ions equilibrate on a timescale much smaller than that of the OPs. Therefore, fluctuations from the solvent modulate the residuals generated within the MD part of the constant OP sampling, and hence affect the thermal-averaged force. If slow hydrodynamic modes are found to be of interest, e.g., to account for long wavelength fluctuations of the solvent during vortex formation or other inertial effects, these atoms can be included in the definition of the OPs. The emergence of such coupled slow modes is also indicated by the appearance of long-time tails in the OP velocity autocorrelation functions. However, such tails are not observed in the simulation study of Sect. V, as is also confirmed via agreement with MD results. When ions are tightly bound to the macromolecule, they are considered a part of the OPs. After every Langevin time step an ion accessible surface is constructed via VMD and ions close to the surface are tracked during the MD ensemble enrichment calculation. Those with appreciable residence time within the surface are included in the definition of the OPs henceforth. A similar solvation scheme has already been utilized with OPs in simulating virus capsid expansion in $Na^+$ and $Ca^{2+}$ solutions [28].

Constructing atomic structures with modest to high Boltzmann probability, that are consistent with the free energy minimizing pathway of the assembly, is often not possible if only subsystem CMs and coarser-grained variables are known. This is because, there are too many structures consistent with the same overall description, only a few of which contribute to the free energy minimizing pathway. Thus, though the above multiscale methodology formally derives Langevin equations from the *N*-atom Liouville equation, it is impractical to apply it as a simulation tool. To overcome this issue, each subsystem is described by a set of subsystem-centered variables that characterize not only their position, but also orientation and overall deformation. The number of all-atom structures consistent with this information is much less than those constrained only by the CM information. Thus, limited (though still quite large) ensemble sizes suffice for average calculations. In the next section, we use conventional MD simulations together with the above mentioned procedures for calculating OPs, thermal forces and diffusions to elucidate scenarios where a dynamical reference configuration is required for capturing assembly dynamics. Finally, we modify the above formalism to address the issue of all-atom reconstruction, and derive a computationally feasible workflow for implementing the current development.

## V. Results and Discussion
### A. Demonstration Systems and Simulations
The multiscale analysis developed in Sect. III yields a Smoluchowski equation for evolving the reduced probability of the OPs and CMs. For practical simulations, Langevin equations were derived from these Smoluchowski equations in Sect. IV, wherein the forces and friction/diffusion coefficients can be obtained via ensemble methods and short MD simulations. An OP-based Langevin simulation algorithm has been developed and implemented as DMS [29-30]. However, in the latter study, OPs are defined in terms of a fixed reference configuration (not a dynamical one as in Sect. II), and any structural change is considered a deformation of this reference structure. In contrast, here, we present a dynamical reference configuration to construct OPs, and subsequently use these OPs to probe the structure and dynamics of a macromolecular assembly. Then, using all-atom data (positions, velocities, and forces) from MD (NAMD [52]) trajectories with classical force fields (CHARMM27 [53]), the behavior of the thermal-average forces and diffusion factors (in the Langevin equations for coupled OP and CM dynamics) are analyzed and compared between contrasting simulations of connected versus disconnected systems. Finally, we deduce the effect of simultaneous $\underline{R}$-$\underline{\Phi}$ Langevin evolution on the accuracy of multiscale macromolecular assembly simulations via their direct comparison with MD predictions. These ideas are demonstrated using MD and multiscale simulations of the RNA mediated assembly of STMV capsid proteins, and the expansion of its capsid-free RNA. This choice of demonstration system is made according to the following criteria: (a) the system must be large enough so that the timescale separation between individual atomistic and overall structural dynamics warrants a multiscale approach; (b) the system must be small enough so that complex dynamical behaviors are observed within 10ns of MD simulation. Also, simulations presented here involving the structural components of STMV are a natural extension of our earlier studies of the same system [29-30]. All simulation parameters are provided in Table I.

      Consider a 10ns MD simulation of STMV protein monomers assembling with RNA in 0.25M NaCl. Initial configuration of the system is a random-coil state of the 949 nucleotide RNA surrounded by 60 randomly placed capsid monomers (accounting for 12 pentamers). During the simulation, proteins become electrostatically attracted towards the RNA and the system starts organizing into an RNA core with an external protein shell. With this, the protein monomers gradually transition from disconnected to a non-covalently bonded state. There exists multiple pathways that lead to such self-assembly behavior in STMV [54]. However, the aim of this study is not to analyze these mechanisms (Fig. 2). Here, it is understood how such a structural transition can be captured by the set of coupled Langevin equations in Sect. IV. To probe contributions from different terms in the Langevin equation as they vary with the nature of system dynamics, we compare OPs, CMs, thermal forces, and diffusions from the MD assembly simulation to those from an RNA expansion in 0.25M NaCl solution. The RNA of STMV is tightly encased within the capsid core in an icosahedral structure via strong electrostatic interactions [54]. As the capsid is removed, electrostatic repulsion among neighboring negatively charged nucleotides causes the system to expand so that the repulsive forces subsides [29]. This simulation provides a contrasting example to that of the assembly as, now, the subunits (pentamers of nucleotide helices (Fig. 2)) are moving further apart and not towards each other. Furthermore, connectivity of the system is maintained throughout the simulation. It is shown how such differences in dynamical behavior are reflected by Langevin evolution of OPs and subsystem CMs.

**B. Timescale of Order Parameter Evolution**

In the following, we examine the slowness in the rate of change of $\underline{\Phi}$ and $\underline{R}$ during an MD simulation. Such analysis validates the timescale separation between slow and fast degrees of freedom that our coarse-grained variables provide. All-atom configurations derived every 100ps from the MD simulation of monomer assembly are used in (II.2) and (II.11) to reconstruct the evolution of CMs and global OPs. The OPs considered (with $k = \{100, 010, 001\}$) capture overall dilation/compression of the STMV-RNA assembly along the three Cartesian directions (Sect. SI1) [29-30], and $\underline{R}$ incorporates CMs of the protein monomers. To investigate how $\underline{R}$ evolution mediates $\underline{\Phi}$ dynamics through changes in $U_k^S$, we introduce the quantity $\Theta_k(t)$ via

$$\Theta_k(t) = \frac{\sum_{S=1}^{N^{sys}} M^S U_k^S(t) U_k^S(0)}{\sum_{S=1}^{N^{sys}} M^S \{U_k^S(0)\}^2} \tag{V.1}$$

where $U_k^S(t) \equiv U_k[\vec{R}^S(t)]$, $\vec{R}^S(t)$ is the position of the $S^{th}$ subsystem CM at time $t$, and $U_k^S(0) \equiv U_k[\vec{R}^S(0)]$. $\Theta_k$ tracks rotations in the basis vectors $U_k^S$ as the reference configuration of CMs $\underline{R}$ changes in time [43]. Smaller changes in $\Theta_k$ suggest the applicability of a fixed reference configuration for $\vec{\Phi}_k$ construction holds for long periods of time.

Time evolution of $\Theta_k$, $\vec{\Phi}_k$ and $\vec{R}^S$ are plotted in Fig. 3 for 10ns. Overall decrease in the magnitude of the dilation-compression OPs and $|\vec{R}^S|$ (i.e., distance of the $S^{th}$ subsystem CM measured from the assembly origin) is consistent with decrease in the assembly radius as the proteins are attracted towards RNA (Fig. 3). These results further imply that the rate of change of polynomials $U$ is slower than that the OPs $\Phi$ which, in turn, is slower than the subsystem CMs $R$. This is because while $U$ and $\Phi$ characterizes motion of the entire system, $R$ implies that for only a subunit. In particular, change in the basis functions $U$ is the slowest as it varies smoothly across the system [19, 43]. Though slow, such changes suggest the use of a dynamical reference configuration for constructing OPs. All three variables, in turn, change on a timescale several orders of magnitude larger than atomic fluctuations. Thus, even though there exists a spatio-temporal scale separation between the three types of coarse-grained variables, it is much smaller in comparison to their separation with atomic scale. As a result, it is assumed that the three variables change on a similar timescale relative to all-atom fluctuations. These evolution characteristics of the coarse-grained variables are consistent with the scale separation arguments made in Sect II that underlies the multiscale development of Sects. III and IV.

## C. Analyzing Thermal Forces and Diffusions from MD Trajectories

The $\vec{\Phi}_k$ and $\vec{R}^S$ dependencies of thermal forces and diffusivities are analyzed to compare their contributions toward capturing overall assembly dynamics. As above, the atomic scale data required for these calculations is extracted at different points in time from the 10ns MD simulations. First, consider the correlations that reflect in diffusion matrix $\bar{\bar{D}}$. An ensemble of all-atom configurations consistent with the instantaneous $\underline{R}$ and $\underline{\Phi}$ state of the MD derived structure is constructed via procedures discussed in Sect. SI2. This ensemble, together with an MC averaging scheme yields the diffusion factors. In Fig. 4(a)-(c) the $\bar{\bar{D}}_{ij}^{RR}$ (III.20), $\bar{\bar{D}}_{kk'}^{\Phi\Phi}$ (III.21) and $\bar{\bar{D}}_{ik}^{\Phi R}$ (III.22) terms of the diffusion matrix are presented, where $k_1 + k_2 + k_3 \leq 6$ implying $3^3$ OPs (II.3) and $S = 1, L\ 60$. The number of OPs results from mass weighted minimization of the residuals (II.6). Both the $\bar{\bar{D}}_{kk'}^{\Phi\Phi}$ and $\bar{\bar{D}}_{ij}^{RR}$ diffusions are strictly diagonally dominant. This is attributed to the use of an orthonormal basis function $U$. By definition, the diagonal terms are positive. Off-diagonal (inter-OP) correlations can both be positive or negative depending on whether the OPs capture motions in same or opposite direction. In analogy to the off-diagonal terms, the $\vec{\Phi}_k$-$\vec{R}^S$ correlations are much smaller than the diagonal $\vec{\Phi}_k$-$\vec{\Phi}_{k'}$, $\vec{R}^S$-$\vec{R}^{S'}$ ones. Each of the three types of correlations grows with time. This is because subsystem motions become more coordinated as they approach one another. Consequently, relaxation time of the OP or CM velocity autocorrelation function increases (Fig. 4(e)-(g)). The $\vec{\Phi}_k$-$\vec{\Phi}_{k'}$ coupling grows the most with time. Since OPs are a function of CMs, they change more slowly and therefore relax on a longer timescale than the latter (Figs. 4(d) and SI2). Thus, as the components become strongly interacting, the diffusion coefficients become strictly ordered $\bar{\bar{D}}_{kk}^{\Phi\Phi} \geq \bar{\bar{D}}_{ii}^{RR} > \bar{\bar{D}}_{ik}^{\Phi R}$.

Now consider the expansion of the free RNA of STMV in NaCl solution. Each of the 12 RNA pentamers (Fig. 2) is considered a subsystem. The diagonal dominance of $\bar{\bar{D}}_{kk}^{\Phi\Phi}$ and $\bar{\bar{D}}_{ii}^{RR}$ is preserved, and cross-terms $\bar{\bar{D}}_{ik}^{\Phi R}$ are still small (Fig. SI3). However, unlike the assembly example, there is a larger difference between the higher and lower-order contributions to $\bar{\bar{D}}_{kk}^{\Phi\Phi}$. This result is explained in Sect. SI3 in terms of differences in OP velocities and relaxation times of their correlation functions, and used later to justify the role of dynamical reference structure in OP evolution. Furthermore, in contrast to the assembly simulation, the three types of correlations decrease with time as expansion causes the RNA to become less coordinated (Fig. 5). However, the contribution of $\bar{\bar{D}}_{ik}^{\Phi R}$ to the overall diffusion matrix is more significant for the RNA expansion than for the protein assembly. This is because the local and global deformations in RNA are correlated through covalent interactions between subunits, whereas such bonds are absent between the aggregating monomers, thereby making local and overall motions less coupled. In both cases $\bar{\bar{D}}_{ik}^{\Phi R}$ increases with higher values of $k$, implying that motion of the higher-order OPs is more strongly correlated to the CMs than the lower-order ones (Figs. 4(h)-(i) and 5(c)-(d)).

Next the thermal forces $\bar{f}_k^{\Phi}$ and $\bar{f}_i^R$ are computed using (III.16) and compared for the two simulations. Given all-atom structures from the MD trajectory, OP-constrained ensembles are constructed, and an MC-based averaging of atomic forces from this ensemble yields thermal forces (Sect. SI2). For both examples $\left|\bar{f}_i^R\right| > \left|\bar{f}_k^{\Phi}\right|$, though the difference is greater for the assembly case. This is expected as OP forces are averages over all subsystems and therefore result in more cancellation than subsystem center forces. The subsystem forces on the free RNA are repulsive while those on the proteins are attractive (Fig. 6). Furthermore, larger forces on the protein monomers versus on the free RNA imply a net free-energy minimizing force when protein-RNA attraction overcomes the intra-RNA repulsion, thereby favoring formation of condensed aggregates.

Combining the MD derived thermal forces and diffusion coefficients within our Langevin framework (IV.2), we deduce the following. Numerically speaking, for the expanding RNA simulation $\bar{f}_i^R > \bar{f}_k^{\Phi}$, however, for the protein assembly we find $\bar{f}_i^R >> \bar{f}_k^{\Phi}$. Furthermore, for the RNA $\bar{\bar{D}}_{ij}^{RR} < \bar{\bar{D}}_{kk'}^{\Phi\Phi}$; whereas for the assembly $\bar{\bar{D}}_{ij}^{RR} \approx \bar{\bar{D}}_{kk'}^{\Phi\Phi}$ (Figs. 4 and 5). With this, additional terms that are included in the Langevin equations to account for a changing reference structure become more relevant for simulating the assembly of proteins than for RNA expansion as summarized in Fig. 7. Physically, these results imply that when a system transitions from a disconnected to connected state, the final structure cannot be perceived as a simple deformation of the initial configuration. This is because there exist intermediate scale forces and correlations that affect the structure and orientation of individual disconnected subsystems, which cannot be accounted for via larger scale factors that imply only global deformation of the assembly. Defining a dynamical reference configuration in terms of subsystem CMs accounts for these intermediate scale changes. Larger scale deformations of this reference structure are mediated through the renewal of polynomials consistent with the evolved reference configuration, construction, and Langevin dynamics of the resulting OPs. With this, the hierarchical organization of an assembly is reflected within our formalism. Use of multiple reference configurations to account for exchange correlation has been very successful in electron structure theory [55]. In our development, multiple reference structures and associated basis functions also add to the range of correlations that the multiscale analysis can capture.

## D. Comparison of Multiscale and MD Trajectories

In the demonstrations of Sects. V.A-C, atomistic configurations for computing forces and diffusions were derived from MD simulations. To implement the multiscale algorithm of Sect. IV as a practical simulation tool, one needs a procedure for constructing ensembles of atomic configurations consistent with the coarse-grained state of the assembly. Such states are characterized by the instantaneous values of global OPs and subsystem CMs. Also, the forces and diffusions to be computed from these ensembles drive the Langevin $\underline{\Phi}$ and $\underline{R}$ dynamics, thereby simultaneously capturing large-scale deformations of the assembly. This way, large, intermediate, and small space-time scale features of the assembly structure and dynamics are precisely represented. However, as mentioned in Sect. IV, too many structures exist that are consistent with the same set of CMs, and only a few of these contribute to the free energy minimizing pathway. For example, complex dynamics such as those of RNA during the assembly simulation cannot be captured only by tracking CMs, and therefore is not included in the above demonstration of hierarchical OP dynamics. To address such issues, our dynamical reference algorithm must include additional intermediate scale variables, denoted $\vec{\varphi}_k^S$, that account for the shape and orientation of subsystems, rather than just those that describe the position of their CM. A general framework for introducing such variables has already been introduced [43] and is mentioned in Sect. SI2. As defined there, $\vec{\varphi}_0^S \equiv \vec{R}^S$ implies subsystem CMs and those with $k = \{100, 010, 001\}$ characterize overall extension-contraction-rotation of the $S^{th}$ subsystem. With this in place, a more descriptive set of coarse-grained variables is achieved for capturing assembly dynamics, and limited sampling is required to obtain atomic configurations that are consistent with the well-defined overall state of the system. Since the subsystem orientation variables change on a timescale similar to that of their CMs (Fig. 3)[43], they satisfy Langevin equations of the same form as (IV.2). Hence the formalism is unchanged by this addition, with the exception of an additional Langevin equation for each new subsystem variable.

The multiscale formalism augmented as above is preliminarily implemented according to the workflow of Fig. 8. To gauge the accuracy of these simulations, their results are benchmarked against those from conventional MD. The RNA mediated protein assembly and expansion of the capsid-free RNA are simulated for 10ns. The multiscale and MD simulations are implemented with identical initial structure and conditions (Sect. V.A).

First consider results from the assembly simulation. The system is described using $3^3$ global OPs ($\vec{\Phi}_k; k_1 + k_2 + k_3 \leq 6$) along with $2^3$ subsystem OPs for each of the 60 monomers ($\vec{\varphi}_k^S; k_1 + k_2 + k_3 \leq 2$), and $3^3$ $\vec{\varphi}_k^S$ variables elucidating the coarse-grained structure of the RNA. The number of these OPs is a natural consequence of residual minimization (II.5-7), and therefore implies maximum structural information at the coarse-grained level (Sect. II). This set of OPs can be systematically enriched if found to be incomplete, i.e., when the OP velocity autocorrelation functions possess long time tails (Sect. IV). Simulation results imply the global, as well as subsystem, thermal-averaged force distribution and diffusion coefficients show excellent agreement with those from MD (Fig. 9(a)-(b)). As above, forces on the monomers are primarily negative implying attractive interaction with the RNA. Such forces facilitate the observed aggregation. Evolution of large-scale structural variables including global and subsystem OPs, radius of gyration, and Root mean square deviation (RMSD) from the initial configuration are presented in Fig. 9(c)-(f). Fig. 9(g) shows the potential energy for the multiscale and MD simulations. These structural variables and energy profiles show excellent agreement in trend as well as in magnitude. As the protein monomers and the RNA aggregate, the potential energy gradually decreases indicating stabilization of the system. This trend is consistent with an increase in the number of inter-nucleic acid hydrogen bonds and suggests that the RNA gains secondary structure during assembly (Fig. SI4). The observed difference is within limits of those from multiple MD runs beginning from the same initial structure with different initial velocities. Agreement in simulated trends, as also visually confirmed in Fig. 9(inset), suggests that the multiscale procedure generates configurations consistent with the overall structural changes that arise in MD. However, care should be taken in comparing atomic scale details, such as dihedral angle or bond length distribution, between the conventional and multiscale simulation procedures. The latter evolves an ensemble of all-atom configurations, e.g., ensembles of size ~$2 \times 10^3$ generated during every Langevin timestep of 50ps, to compute forces and diffusions. The thermal-averaged forces remain practically unchanged with further increase in ensemble size. Thus, such trajectories should be compared to an average synthetic of multiple MD (or a single very long MD) simulations. Such long MD simulations are not practical for the present system and are therefore avoided, though equivalence of our multiscale simulations with ensemble MD methods at the atomic scale has been thoroughly investigated for smaller systems such as lactoferrin [51, 56].

The MD and multiscale trajectories of RNA expansion also show comparable trends. This data is not presented here for the sake of brevity. However, the agreement in simulation results is expected as the RNA of STMV has been very well characterized by our OP-based multiscale method, DMS, in earlier studies involving DMS-MD simulation comparison [29]. Since the methodology of Sects. II-IV is identical to that underlying DMS when the reference configuration is held fixed, and terms (forces and diffusions) in the Langevin equations responsible for evolving the reference configuration are found to be less important for the isolated RNA simulation (Fig. 7(c)-(d)), our present results are equivalent to the earlier DMS ones which in turn, showed excellent agreement with MD predictions [29]. Next, multiscale trajectories for both the demonstration systems are repeated using a fixed versus dynamical reference structure. In Fig. 9(f) the resulting RMSDs from identical initial structures are presented. For the RNA, the multiscale trajectories with and without re-referencing show considerable agreement with those from MD. In contrast, for the protein assembly, only the trajectory with dynamical reference structure shows agreement with MD. From this result, independent multiscale simulations based on the methodology of Sect. IV and workflow of Fig. 8 further confirm the need for the coupled evolution of reference configuration CMs, OPs and atomic ensembles to account for complex deformations in macromolecular assemblies. This MD-multiscale simulation comparison supports the ideas presented in Fig. 7, providing another consistency-check to our methodology and results.

## VI. Conclusions

The present study provides a self-consistent theory of soft matter wherein a set of equations is derived to describe the coupled stochastic dynamics of OPs and molecules constituting a soft material. It also introduces an algorithm to co-evolve OPs and individual molecules over long periods of time so as to enable a calibration-free theory of soft matter. More precisely, the set of OPs and associated multiscale algorithm incorporates the hierarchical nature of soft matter assembly architecture. In contrast with our previous formulation [19, 42], the hierarchical OPs bear a clearer physical interpretation as each of the structural subsystems are considered as separate entities with internal structure. Overall subsystem deformations like extension, compression, rotation, and translation, as well as resulting inter-system motions that probe the temporal dynamics of the assembly are accounted for via this new formulation. In addition, rapid motions such as internal subsystem dynamics or high frequency fluctuations can be probed via a quasi-equilibrium ensemble of all-atom configurations. Unlike MD, in which timesteps are limited to $10^{-14}$ seconds or less, the present theory allows timesteps that are many orders of magnitude greater. For MD or lumped MD approaches, there is a limitation on the motion of the atoms per timestep because of angstrom-scale overlaps. Contrastingly, the $\underline{\Phi}$ and $\underline{R}$ variables within the present approach represent an ensemble of configurations which are weighted by a Boltzmann factor. Here, the driving force is free energy dynamics. While there is a computational burden from the construction of the thermal-average forces and diffusion factors, the associated CPU time is more than offset by the large Langevin timesteps. In addition, since the velocities of particles which reside far distances from one another are effectively uncorrelated, the need to compute large numbers of velocity autocorrelation functions is drastically reduced, allowing for a large savings in the overall computational burden of such simulations. In summary, we propose a force field-based methodology that takes advantage of the structural hierarchy natural to macromolecular assemblies in defining the system as a collection of mutually interacting subsystems with internal dynamics, which simultaneously preserves the all-atom description.

**Acknowledgements**

We appreciate and acknowledge the support of the NSF CRC program, the NSF Division of Mathematical Sciences (under grants DMS-0908413 and DMS-1211667), NIH, Indiana University's College of Arts and Sciences, and the United States Naval Academy Research Council.

| Parameter | Values |
| --- | --- |
| Temperature | 300K |
| Langevin damping | 5 |
| Timestep | 1fs |
| fullElectFrequency | 2fs |
| nonbondedFreq | 1fs |
| Box size | 160Å x 160Å x 160Å[*] |
| | 250Å x 250Å x 250Å[**] |
| Force-field parameter | par_all27_prot_na.prm |
| 1-4scaling | 1.0 |
| Switchdist | 10.0 Å |
| Cutoff | 12.0 Å |
| Pairlistdist | 20.0 Å |
| Stepspercycle | 2 |
| Rigid bond | Water |

**Table 1.** Input parameters for the NAMD simulations.
[*]RNA expansion [**]Protein-RNA assembly

**Figure Captions**
**Fig. 1** OPs characterizing system-wide features affect the relative probability of the atomistic configurations which, in turn mediates the forces driving OP dynamics. This feedback loop is central to a complete multiscale understanding of soft matter systems and the true nature of their dynamics.

**Fig. 2** **(a)** Initial (0ns) and **(b)** final (10ns) structures of the STMV protein monomers aggregating with RNA; **[(c), (d)]** same as (a) and (b) respectively, but for the expansion of free RNA in aqueous solution. These simulations are chosen to illustrate two contrasting scenarios of inter-subsystem interaction, and their effect on Langevin dynamics of the OPs. Random coil structure of the RNA in (a) is generated using the ROSETTA server, and the icosahedral symmetric form is extracted from its encapsidated state as also used in [9].

**Fig. 3** Time evolution of **(a)** $\Theta_k$, **(b)** $\bar{\Phi}_k$; $k = \{100, 010, 001\}$ and **(c)** $\vec{R}^S$; $S = 1, \cdots 60$ are plotted showing that the rate of change of the basis-functions is less than that of the OPs which, in turn, are slower than the subsystem CMs. **(d)** The decrease in protein-RNA aggregate radius is consistent with the decrease of the extension-compression OPs. Characteristic times associated with these three variables are much greater than those of the atomistic fluctuations (tens of ps versus fs).

**Fig. 4** **[(a)-(c)]** Normalized diffusion matrices $\bar{\bar{D}}_{kk'}^{\Phi\Phi}$, $\bar{\bar{D}}_{ij}^{RR}$ and $\bar{\bar{D}}_{ik}^{\Phi R}$ with $k_1 + k_2 + k_3 \leq 6$ (implying $3^3$ OPs) and $S = 1,...,60$ showing that they are diagonally dominant, and $\bar{\bar{D}}_{kk}^{\Phi\Phi} \geq \bar{\bar{D}}_{ii}^{RR} > \bar{\bar{D}}_{ik}^{\Phi R}$ for the protein monomer RNA aggregation. **(d)** Diagonal diffusion terms of (a) and (b) clearly show the noted trend. Dominant diagonal terms of diffusion matrices. **(e)** $\bar{\bar{D}}_{kk'}^{\Phi\Phi}$ and **(g)** $\bar{\bar{D}}_{ij}^{RR}$, and the entire $\bar{\bar{D}}_{ik}^{\Phi R}$ matrix. **[(h)-(i)]** computed after 1 and 10ns of MD simulation showing that all three types of correlations grow. This suggests large-scale motions of the system become more coordinated with time. The $\bar{\bar{D}}_{kk'}^{\Phi\Phi}$ term increases most as it implies system-wide collective motion. **(f)** Normalized OP velocity autocorrelation functions (VACF) for OP $\Phi_{100X}$, which captures overall extension-compression along the X-direction (refer to SI), at times 1 and 10ns. The relaxation time increases as the system shrinks, thereby suggesting an increase in coherence of collective modes accompanying growth in inter-monomer interaction during the assembly process.

**Fig. 5** Dominant diagonal terms of diffusion matrices **(a)** $\bar{\bar{D}}_{kk'}^{\Phi\Phi}$ and **(b)** $\bar{\bar{D}}_{ij}^{RR}$ and the entire $\bar{\bar{D}}_{ik}^{\Phi R}$ matrix **[(c)-(d)]** at 1 and 10ns displaying the decrease of all three types of correlations during the expansion of free RNA aqueous solution. This suggests large-scale motions of the system become less coordinated with time.

**Fig. 6** Distributions of typical subsystem CM ($\bar{f}^R$) and OP ($\bar{f}^\Phi$) forces extracted from the MD simulations of monomer assembly and the RNA expansion. For the assembly problem, $\bar{f}^R$ (orange-dashed) and $\bar{f}^\Phi$ (red) are primarily negative indicating attractive forces on the pentamers, while for RNA the same forces (indicated using blue-dashed and black respectively) are positive implying repulsion between the negatively charged nucleotides. For both examples, the data suggests $\bar{f}_i^R > \bar{f}_k^\Phi$. In particular, for the assembly we find $\bar{f}_i^R \gg \bar{f}_k^\Phi$.

**Fig. 7** Normalized products of diffusion coefficients and free energy minimizing thermal-averaged forces that appear on the RHS of Langevin-$\Psi$ equations, versus time. Data suggests **(a-b)** during assembly, products involving the subsystem CM forces and coupling of CMs with OPs are much greater than those involving only the OPs. This implies that the free energy minimizing nature of the process requires the CMs, and hence the associated reference configuration, to change. **(c-d)** For the free RNA, the OP forces and diffusions dominate those associated with CMs, implying the free energy minimizing forces require the reference configuration to be kept fixed for longer periods of time than in (a).

**Fig. 8** Workflow illustrating computational implementation of the hierarchical OP-based multiscale algorithm. Boxes indicating computations with all-atom details are presented in magenta, while those involving OPs are shown in yellow.

**Fig. 9** Distributions of typical **(a)** subsystem CM (dotted) and OP ($\Phi_{100X}$) (solid) forces, and **(b)** associated velocity autocorrelation functions from 10ns MD and multiscale simulations of STMV

protein-RNA assembly. Results are consistent with those in Figs. 6 and SI2. **(c)** Evolution of large-scale structural variables including **(c)** global OPs, **(d)** subsystem CMs, **(e)** radius of gyration, and **(f)** RMSD from the initial configuration. All numerical results show excellent agreement between MD and multiscale predictions. RMSD from the initial structure also shows that the Langevin assembly simulation requires referencing to reproduce MD results; however, capturing the RNA expansion does not require a dynamical reference structure. Dashed line implies the use of a dynamical reference structure, whereas dash-dots imply a fixed reference structure during Langevin evolution of the system. **(g)** The potential energy profiles for the multiscale and MD simulations also show strong agreement. (inset) visual comparison of multiscale (blue) and MD (red) generated all-atom structures of the assembled monomers and the RNA. Also provided are positions of subsystem CMs in bead representation (multiscale (black) and MD (red)) showing almost identical configurations.

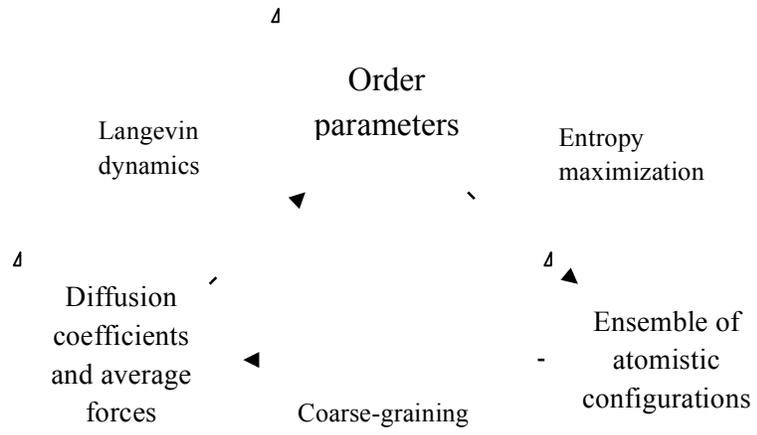

**Fig. 1**

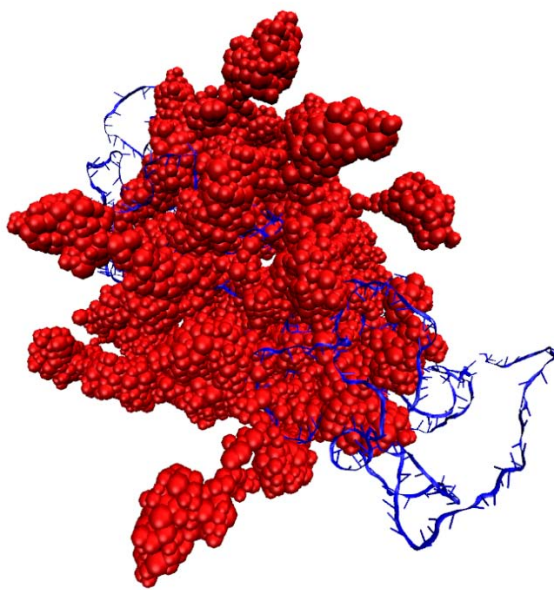
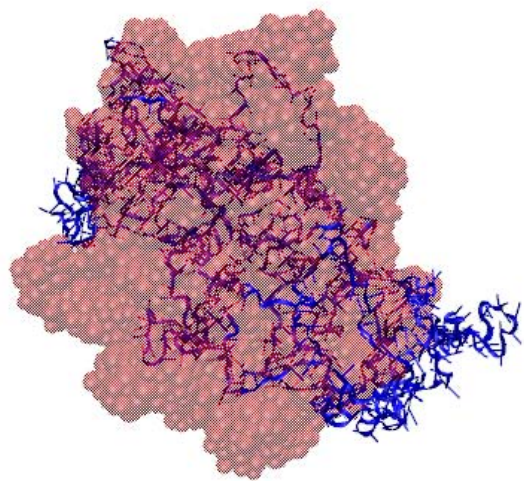

(a)

(b)

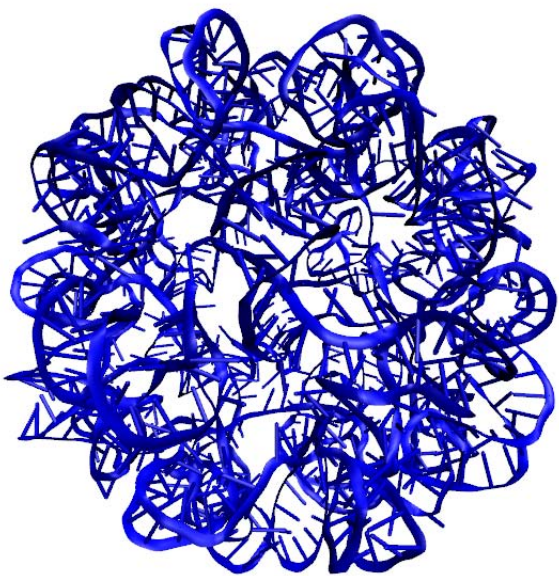
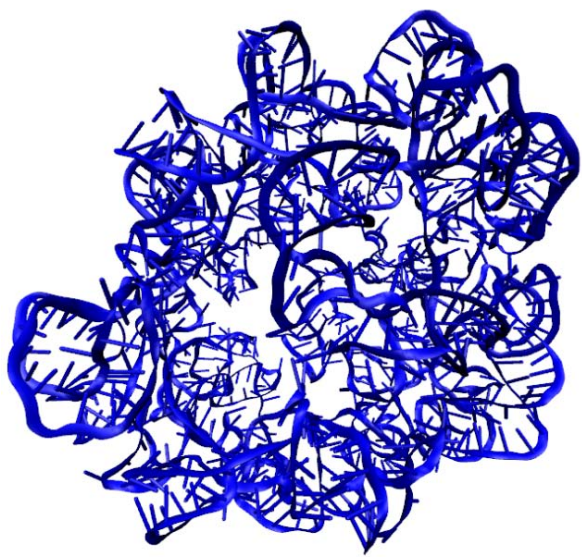

(c)

(d)

**Fig. 2**

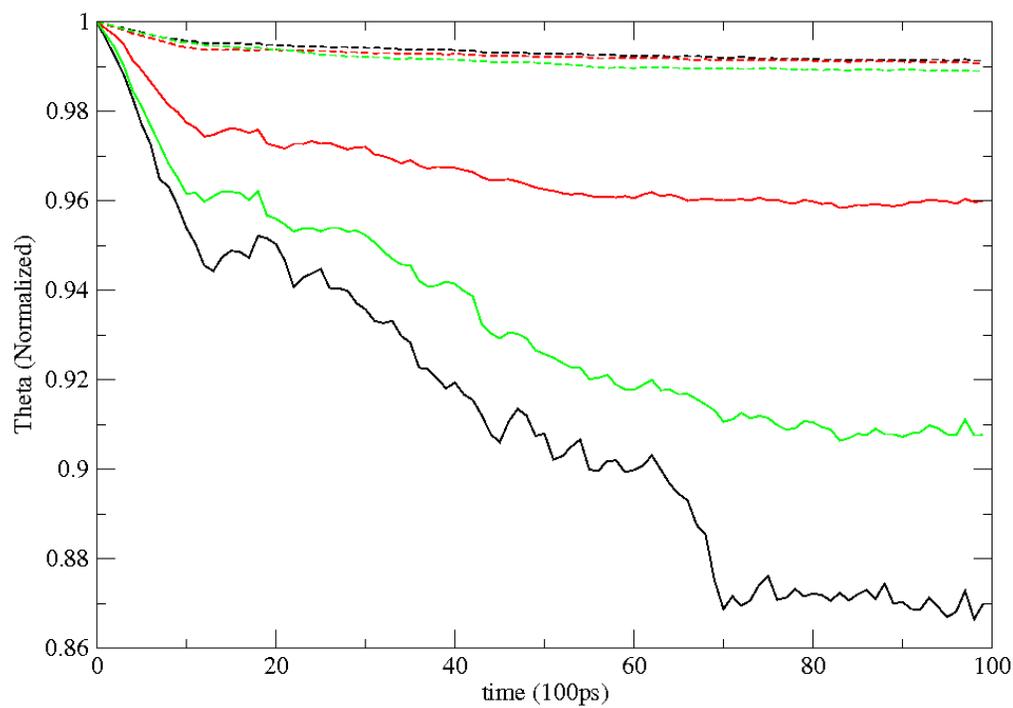

**(a)**

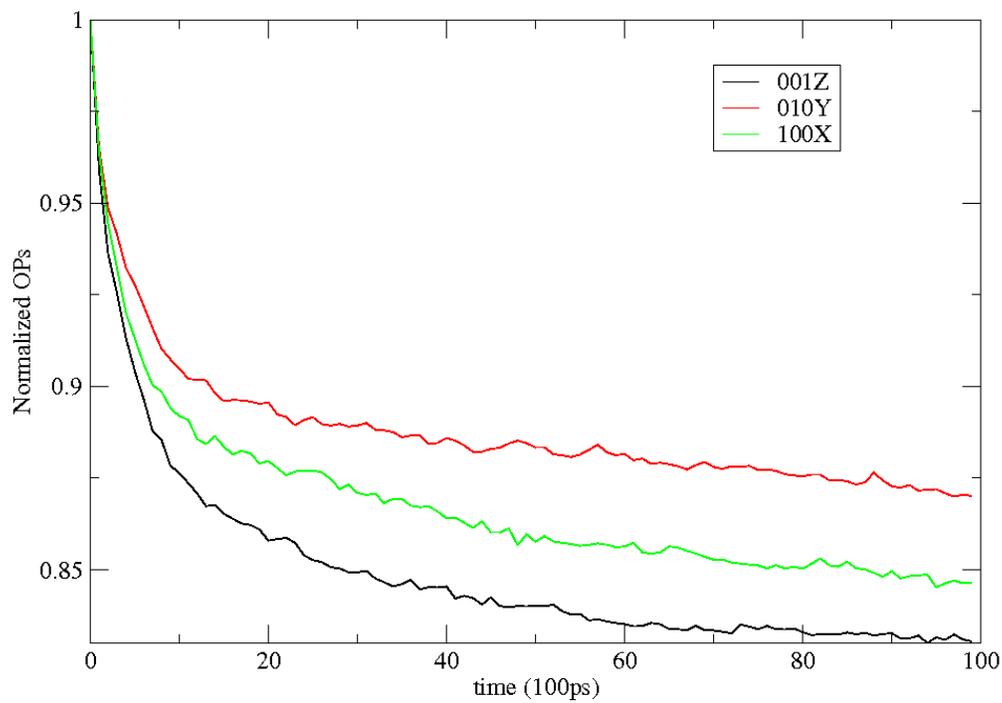

**(b)**

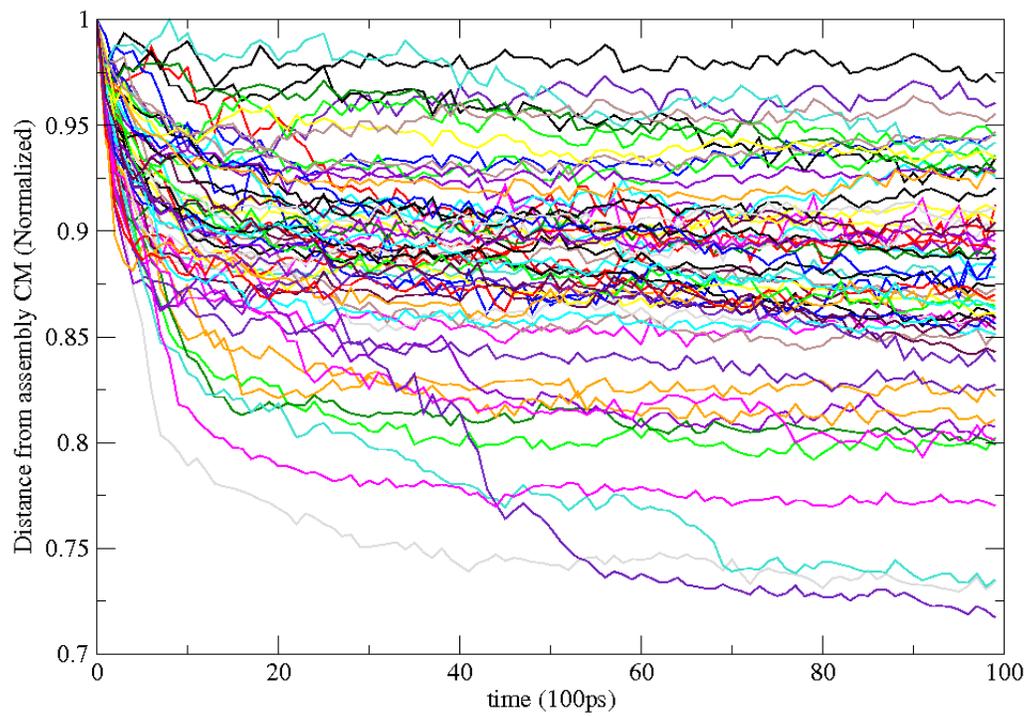

**(c)**

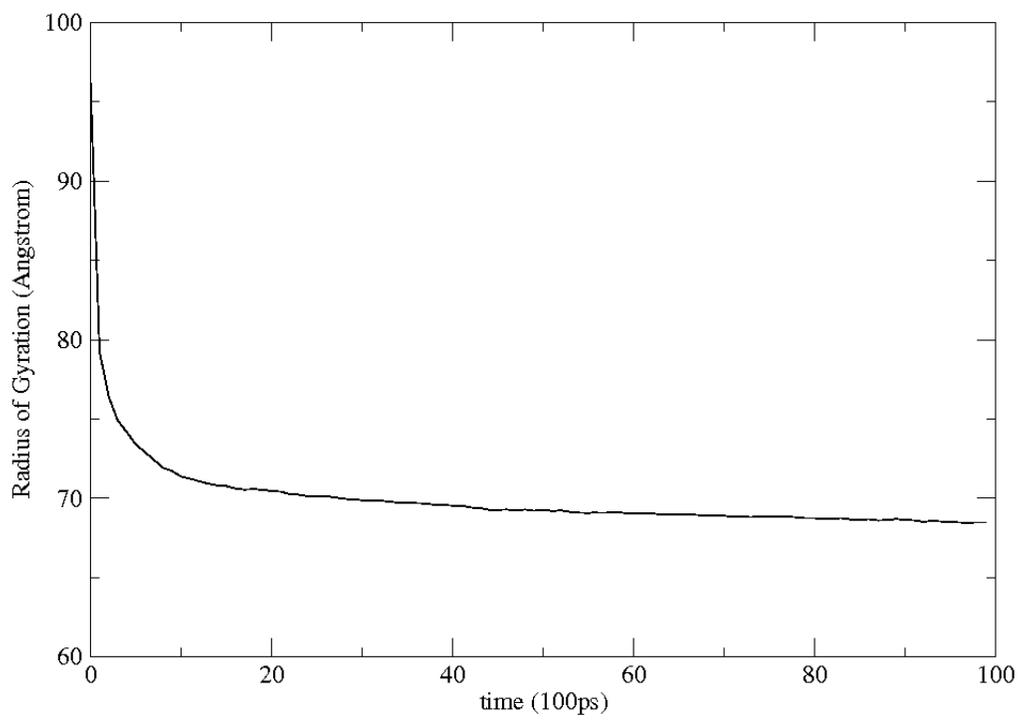

**(d)**

**Fig. 3**

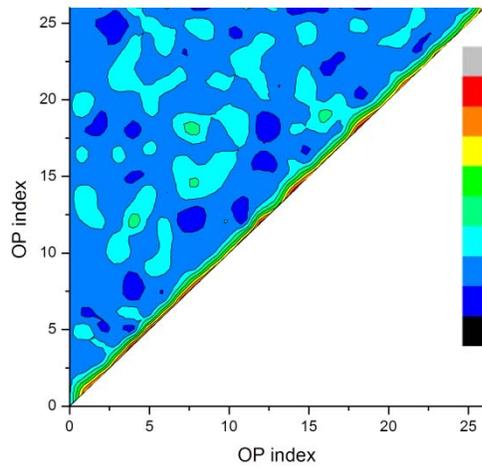
(a)

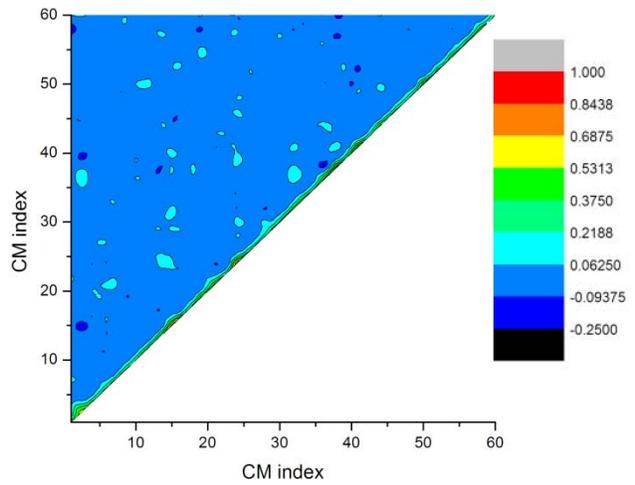
(b)

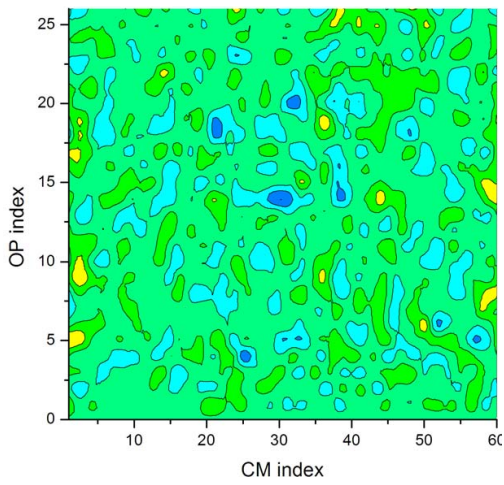
(c)

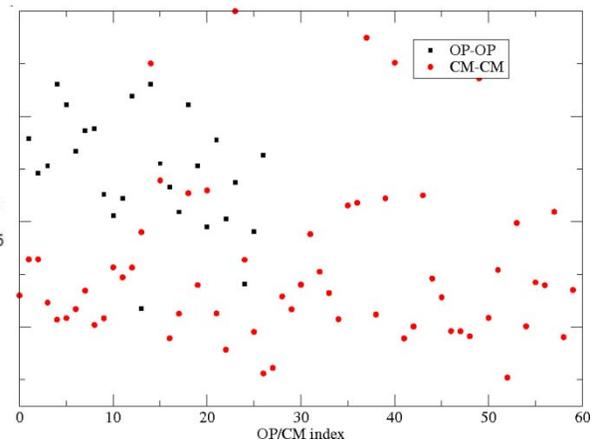
(d)

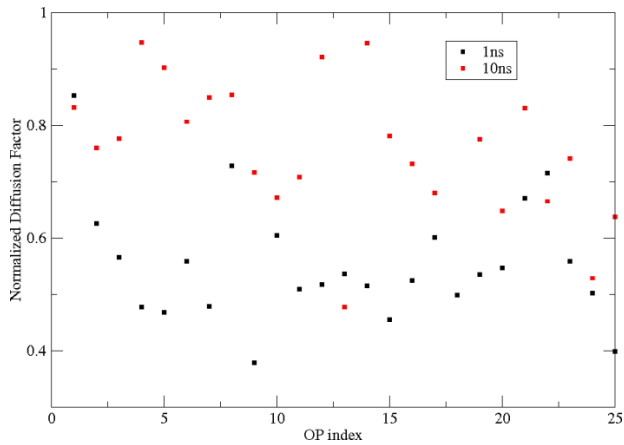
(e)

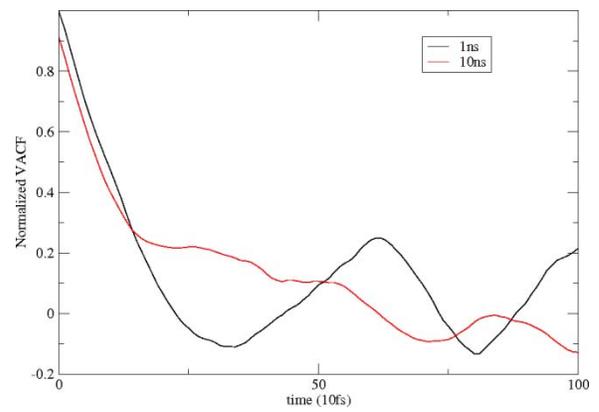
(f)

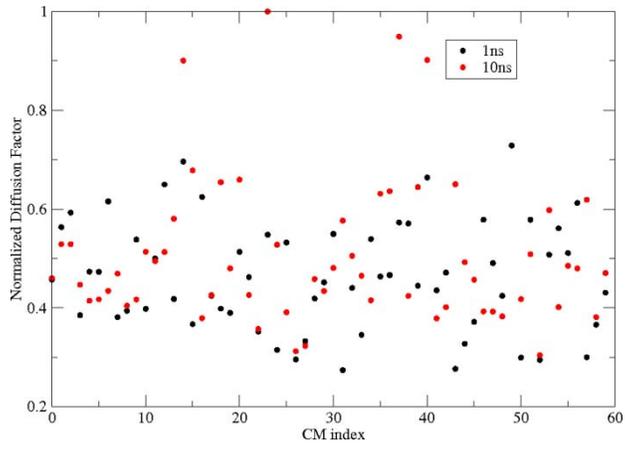

(g)

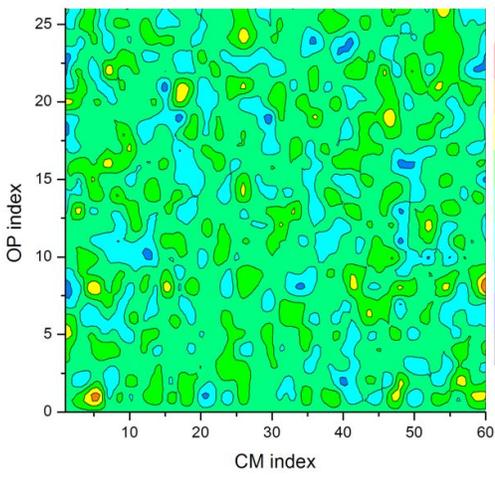

(h)

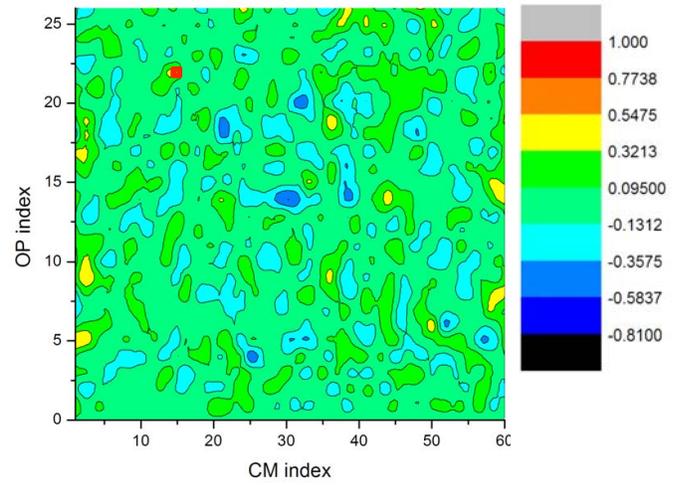

(i)

Fig. 4

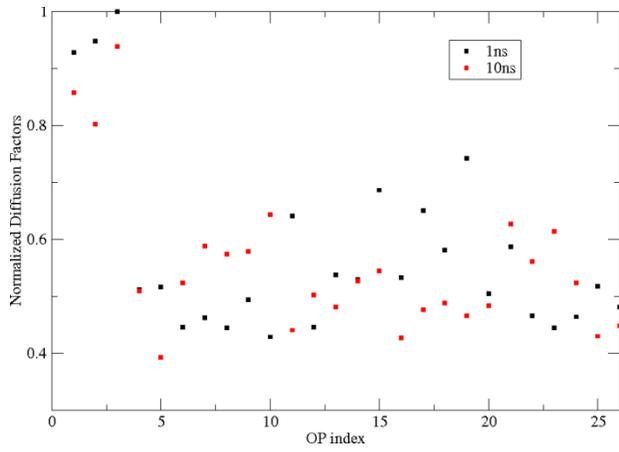
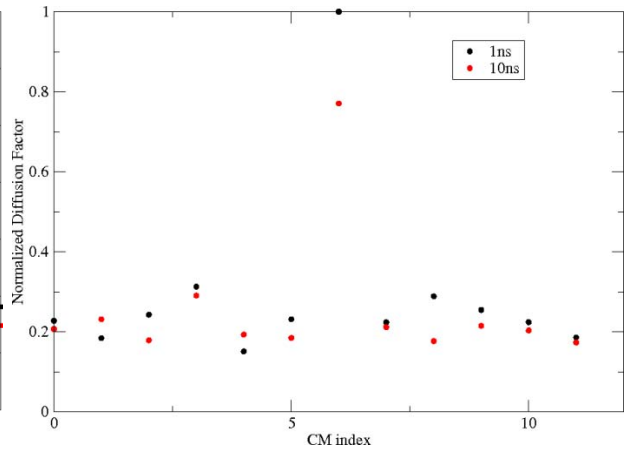
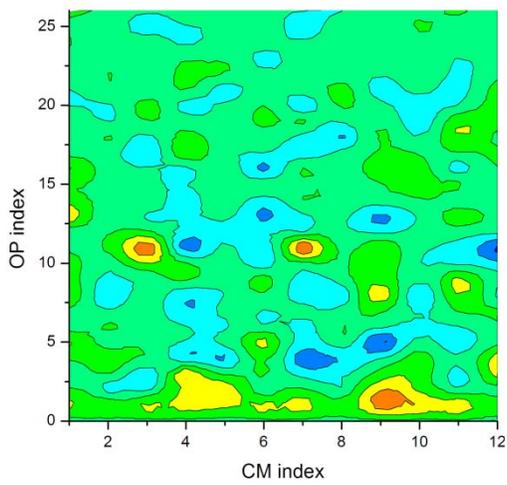
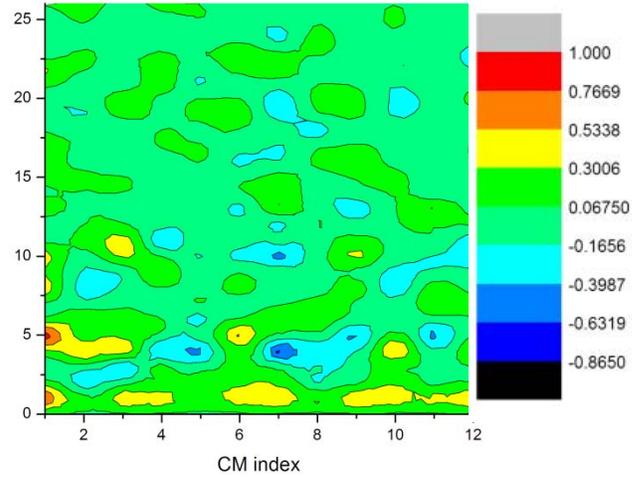

**Fig. 5**

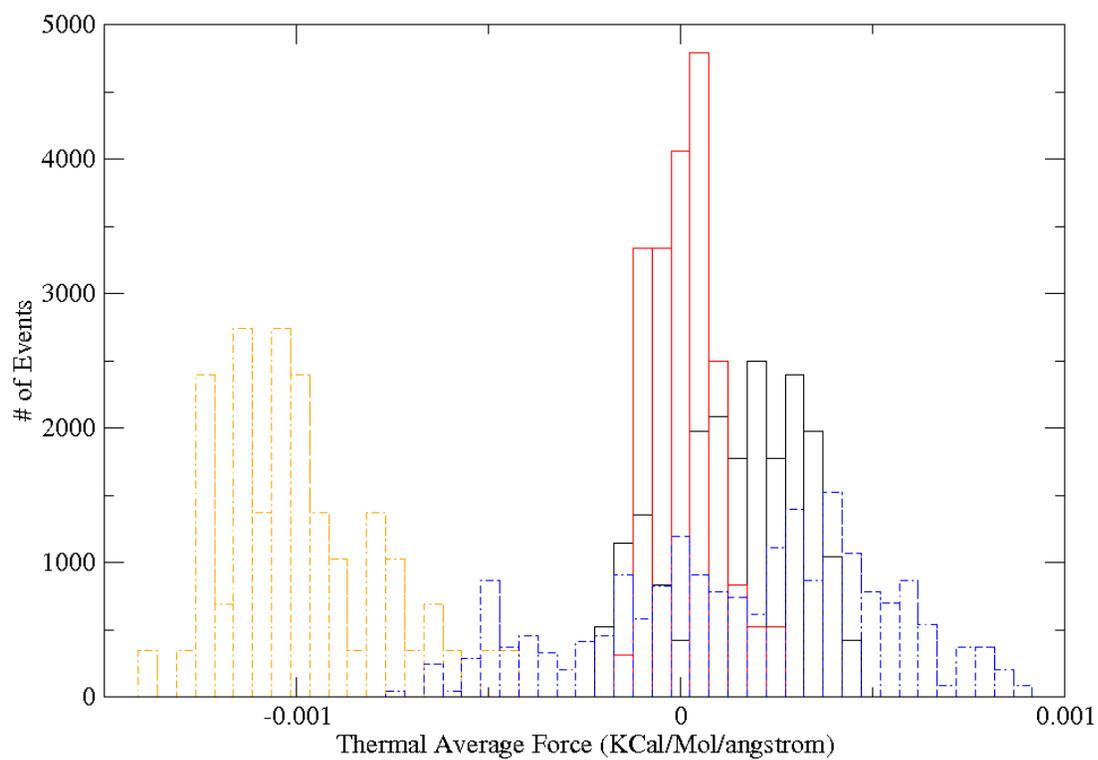

**Fig. 6**

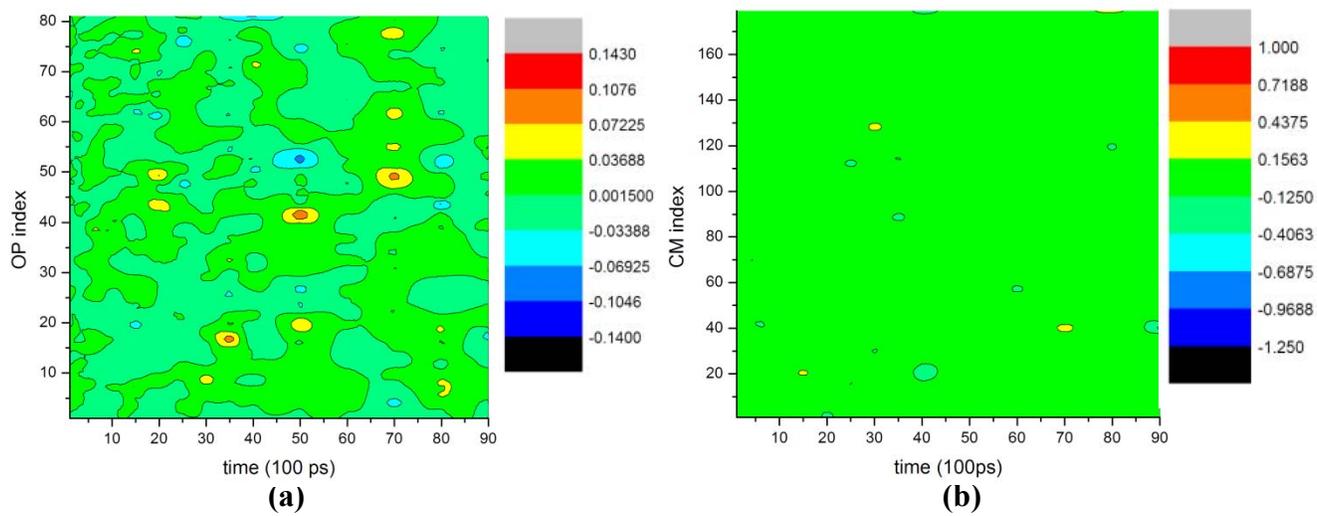

(a)                      (b)

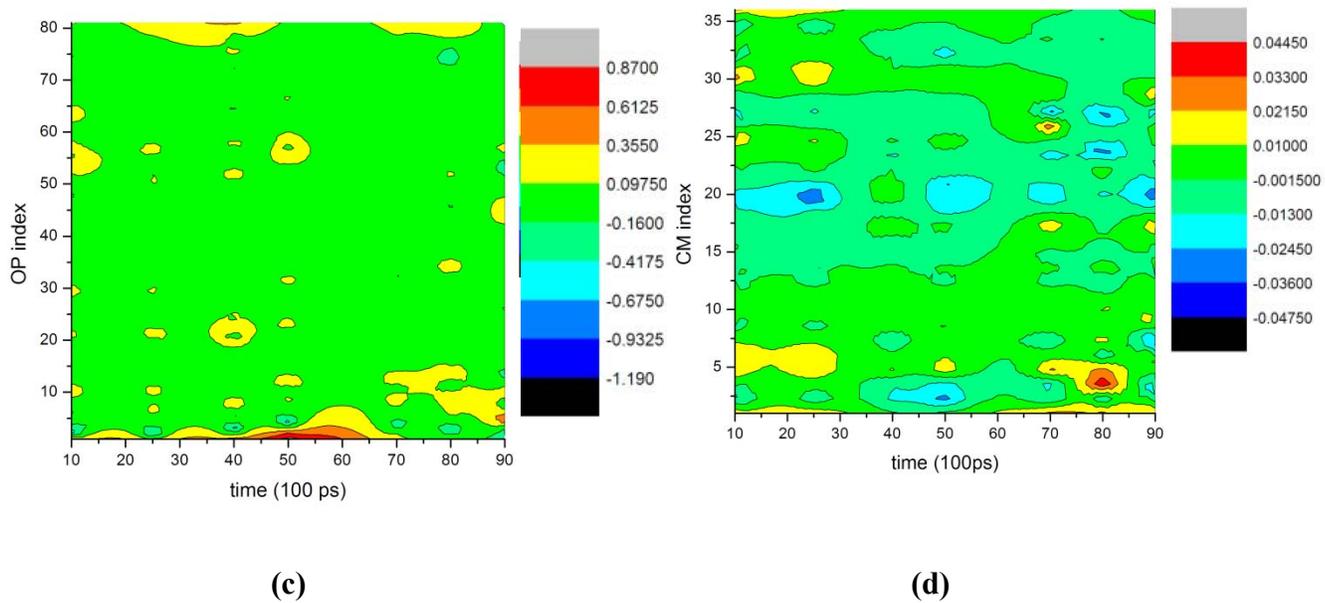

(c)                            (d)

**Fig. 7**

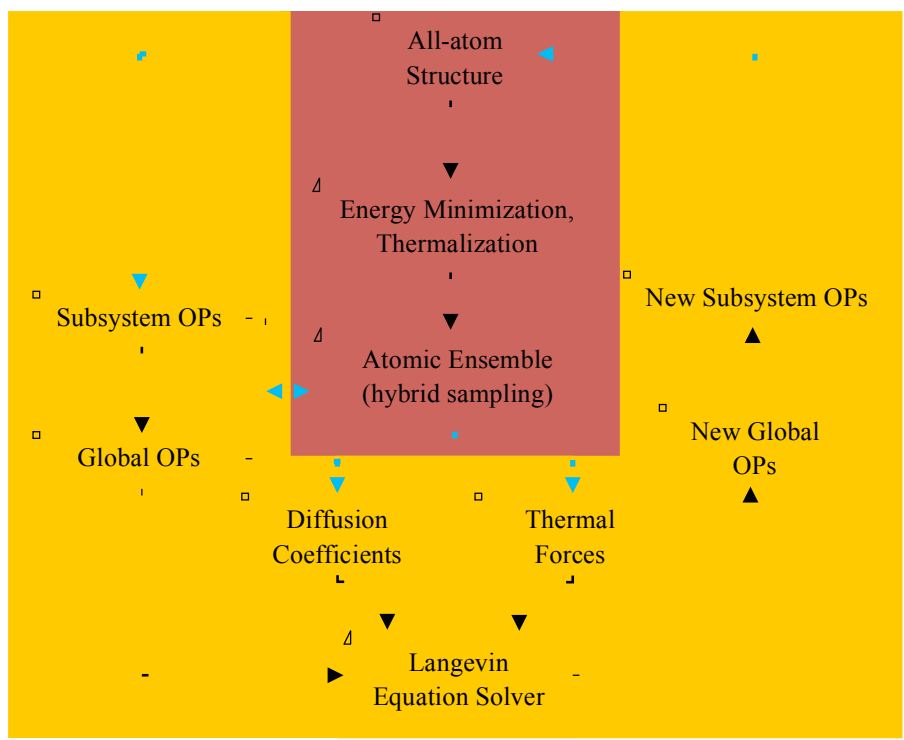

**Fig. 8**

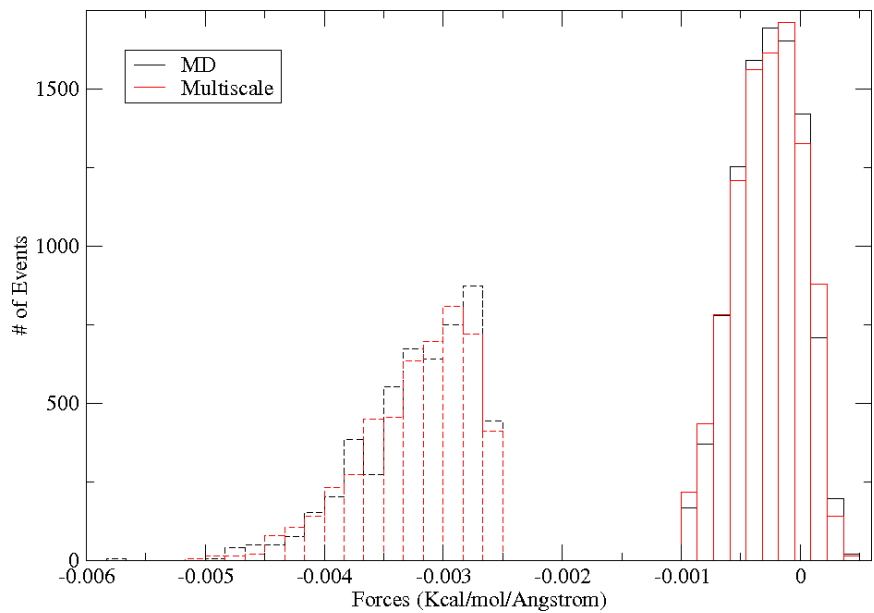

**(a)**

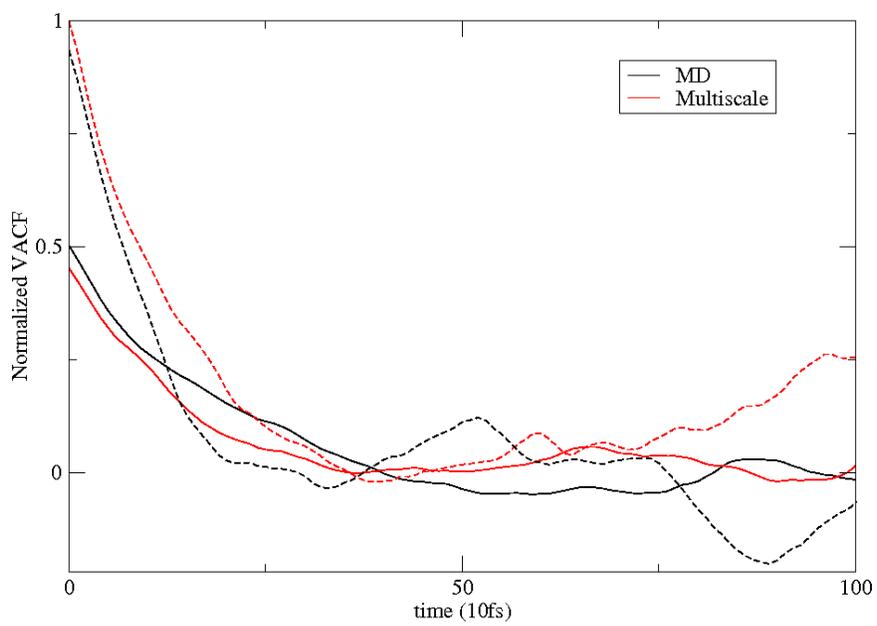

**(b)**

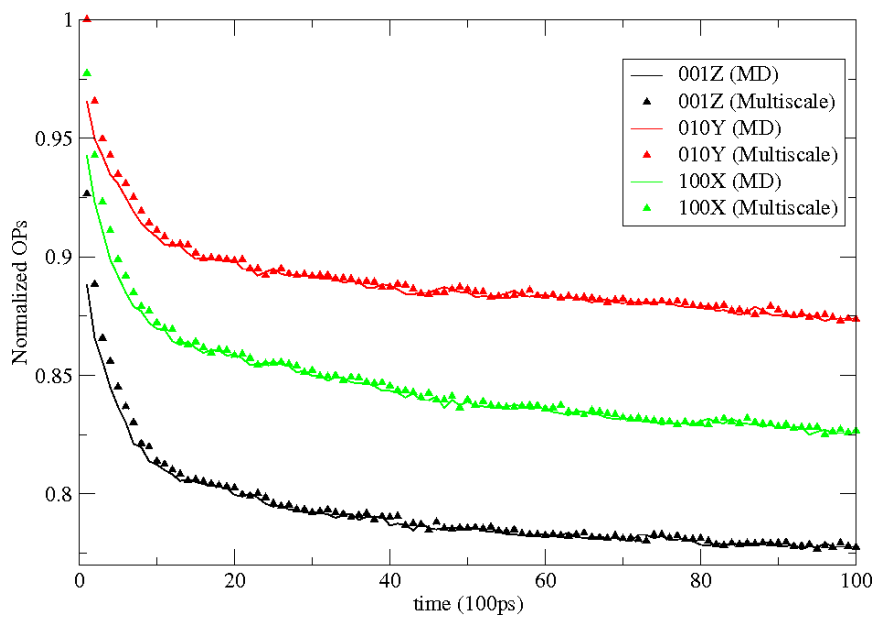

**(c)**

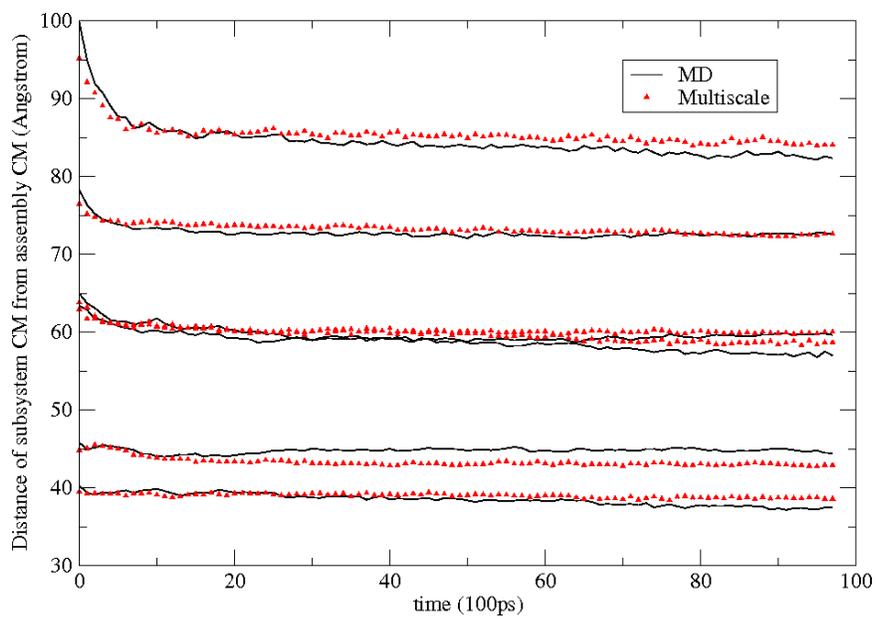

**(d)**

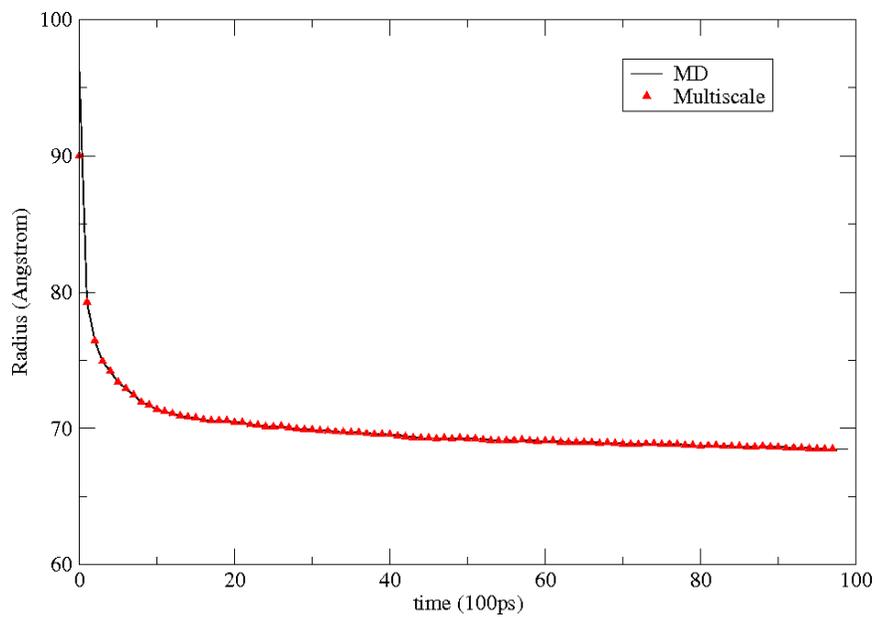

**(e)**

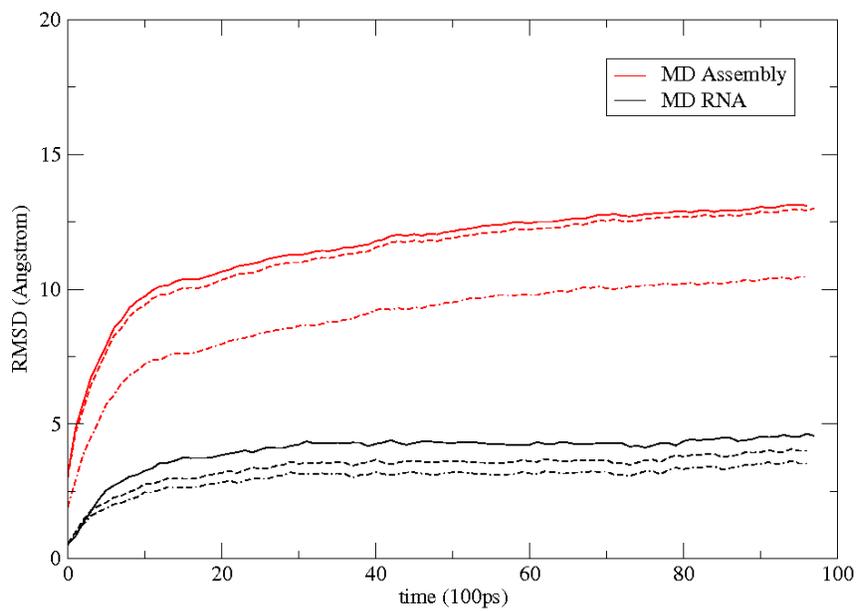

**(f)**

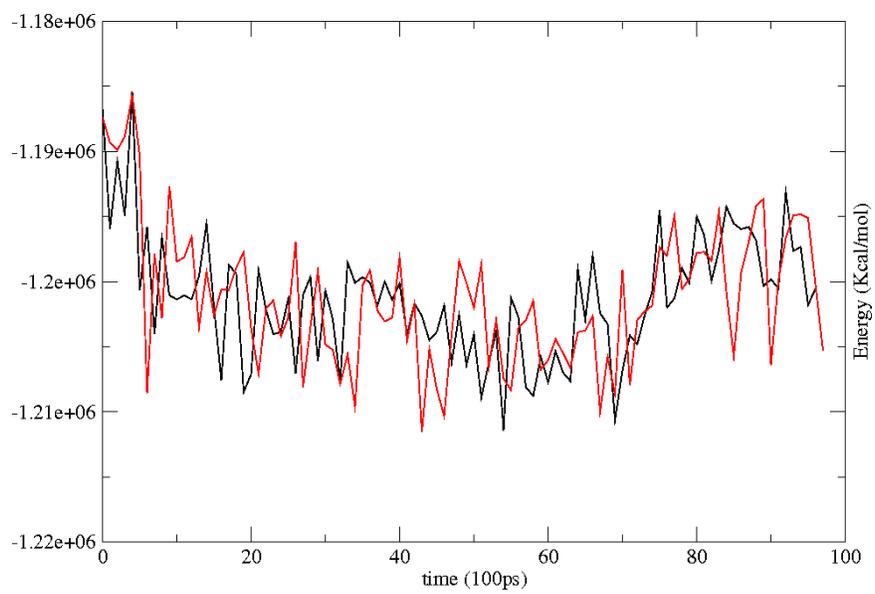

**(g)**

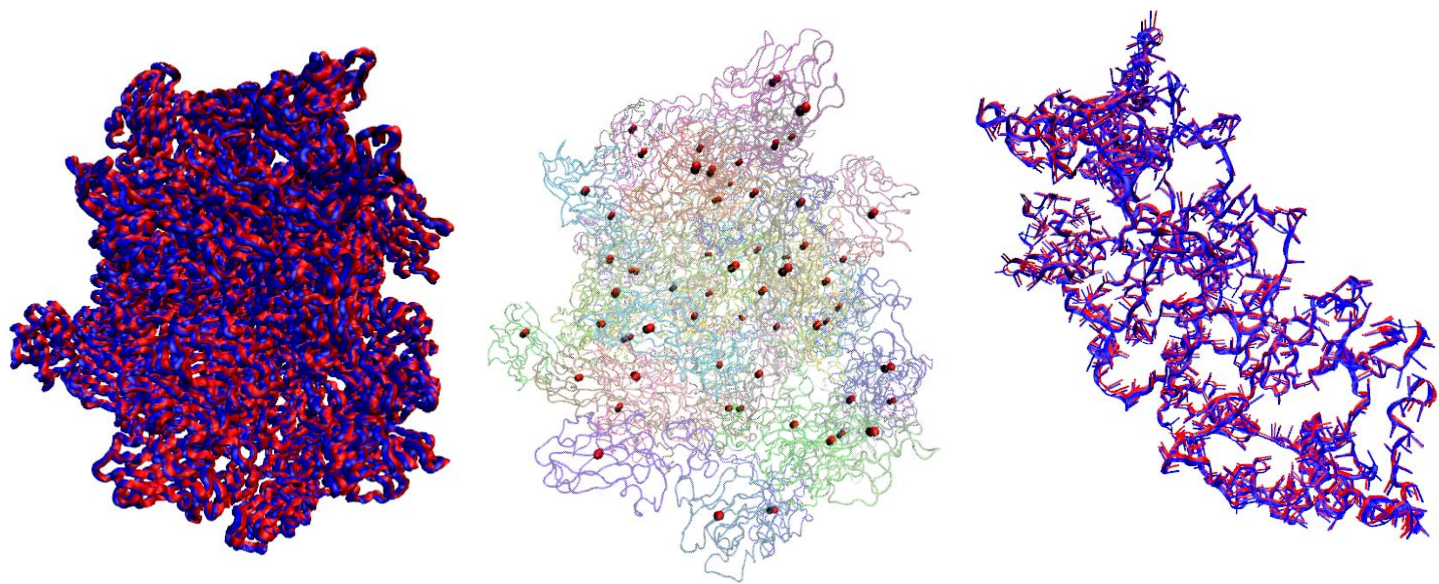

**(inset)**